\documentclass[A_4paper,11pt, final]{article}
\pdfoutput=1
\usepackage[utf8]{inputenc}
\usepackage{jheppub}
\usepackage{tensor}
\usepackage{amsmath}
\usepackage{graphicx}
\usepackage{subcaption}

\usepackage[inline]{showlabels}

\usepackage{hyperref}
\usepackage[capitalize,nameinlink]{cleveref}

\graphicspath{{./figs/}}

\usepackage{xspace}
\newcommand*{\ie}{i.e., }
\newcommand*{\eg}{e.g., }
\newcommand*{\fig}{fig.\@\xspace}

\newcommand*{\eqs}{eqs.\@\xspace}

\title{Inflationary and Gravitational Wave Signatures\newline of Small Primordial Black Holes as Dark Matter}

\author[a,b,c]{Will Barker,}
\author[d,b,c]{Benjamin Gladwyn,}
\author[e,f,g]{Sebastian Zell}

\affiliation[a]{Central European Institute for Cosmolgy and Fundamental Physics, Institute of Physics of the Czech Academy of Sciences, Na Slovance 1999/2, 182 00 Prague 8, Czechia}
\affiliation[b]{Astrophysics Group, Cavendish Laboratory, JJ Thomson Avenue, Cambridge CB3 0HE, UK}
\affiliation[c]{Kavli Institute for Cosmology, Madingley Road, Cambridge CB3 0HA, UK}
\affiliation[d]{Department of Physics, Keble Road, University of Oxford, OX1 3RH}
\affiliation[e]{Arnold Sommerfeld Center, Ludwig-Maximilians-Universit\"at, Theresienstraße 37, 80333 M\"unchen, Germany}
\affiliation[f]{Max-Planck-Institut für Physik, Boltzmannstr. 8, 85748 Garching b.\ M\"unchen, Germany}
\affiliation[g]{Centre for Cosmology, Particle Physics and Phenomenology -- CP3,
	Universit\'e catholique de Louvain, B-1348 Louvain-la-Neuve, Belgium}

\emailAdd{wb263@cam.ac.uk}
\emailAdd{benjamin.gladwyn@physics.ox.ac.uk}
\emailAdd{sebastian.zell@lmu.de}

\abstract{
	Mounting evidence suggests that the semi-classical description of a black hole breaks down at the latest after losing an~$O(1)$ fraction of its mass. As a result, effects such as memory burden can slow down evaporation so that small primordial black holes (PBHs), in particular those in the mass range~$10^{6}\, \text{g}$ to~$10^{9}\, \text{g}$, become viable dark matter candidates. In this paper, we investigate the production of PBHs from a prototype model of polynomial inflation with a non-minimal coupling to gravity. We show that a sufficiently small PBH mass alleviates any tension with CMB observations. Moreover, we develop efficient numerical procedures to identify model parameters and evolve Mukhanov-Sasaki modes to place bounds on the scalar-induced stochastic gravitational wave (GW) background. Whilst we identify some prospects for observation with future GW detectors, our results highlight the need to develop new experiments for high-frequency GW detection in the~$\sim\text{kHz}$ to~$\sim\text{MHz}$ range. Finally, we demonstrate that previously-used ans\"atze for modelling the power spectrum only yield a reliable approximation for the GW signal if some input from inflation is used.
	}

\date{}

\makeatletter
\gdef\@fpheader{\phantom{text}}
\makeatother

\begin{document}
	
	\maketitle
	
	\section{Introduction}
	\label{sec:intro}
	
	Arguably, the longest-standing open question in cosmology concerns the nature of dark matter~\cite{Zwicky:1933gu}. Among the leading candidates are primordial black holes (PBHs) formed in the early Universe due to density fluctuations exceeding gravitational collapse thresholds~\cite{Zeldovich:1967lct,Hawking:1971ei,Carr:1974nx,Chapline:1975ojl} (see~\cite{Carr:2020xqk,Green:2020jor,Escriva:2022duf} for recent reviews). What makes this proposal unique is the fact that the proposed constituents, \ie black holes, are already known to exist in Nature, as has been confirmed through the recent breakthrough-discovery of gravitational waves (GWs) emitted by a merger of binary black holes~\cite{LIGOScientific:2016aoc}.
	
	PBH abundances are constrained by various astrophysical and cosmological observations. There is a growing belief that PBHs of a single mass can only constitute the entirety of dark matter if their mass is in the asteroid range roughly between~$10^{17}\, \text{g}$ and~$10^{22}\, \text{g}$~\cite{Carr:2020gox} (see however~\cite{Carr:2023tpt} for a discussion of possible other mass ranges). One of the most impactful bounds informing this belief comes from the Hawking evaporation of black holes~\cite{Hawking:1974rv,Hawking:1975vcx}, which leads to the requirement that dark matter in the form of PBHs must not have evaporated by today. Computations of the lifetime of PBHs most commonly rely on an assumption of \emph{self-similarity}, implying that an old black hole, one that has already evaporated a significant fraction of its initial mass, is completely indistinguishable from a young, newly formed, black hole of the new mass. If black hole evolution is self-similar in this manner, then Hawking's result for the rate of particle production remains valid throughout the entire lifetime, and it immediately follows that all PBHs of mass below~$10^{14} \, \text{g}$ cannot lead to a viable dark matter candidate since they must have evaporated by the current epoch. This is the widely-assumed \textit{Hawking bound}.

However, the computation of Hawking evaporation is only exact in the semi-classical limit, in which the black hole is eternal and its metric is held fixed. This leads to a question about how long the semi-classical description of a black hole stays valid when its mass shrinks~\cite{Page:1979}. In particular, self-similarity can break down once back-reaction has become strong and the mass has changed appreciably, and consequently particle production may deviate from Hawking's semi-classical result~\cite{Dvali:2011aa,Dvali:2012rt,Dvali:2012en,Dvali:2012wq,Flassig:2012re,Dvali:2013vxa,Dvali:2013eja}. It is important to emphasize that this phenomenon is not connected to large curvature -- instead crucial quantum properties arise for a macroscopic object~\cite{Dvali:2011aa,Dvali:2012rt,Dvali:2012en,Dvali:2012wq,Flassig:2012re,Dvali:2013vxa,Dvali:2013eja}.
 
Indeed, there is mounting evidence that the semi-classical description ceases to be valid at the latest after a black hole has lost half of its mass, both because of entanglement between the evaporation products~\cite{Page:1993wv} and with the inside of the black hole~\cite{Dvali:2013eja}, as well as due to \textit{memory burden}~\cite{Dvali:2018xpy,Dvali:2018ytn,Dvali:2020wft}.\footnote
{Analogous results hold for de Sitter and inflationary spacetimes~\cite{Dvali:2013eja,Berezhiani:2016grw,Dvali:2017eba,Dvali:2020etd,Berezhiani:2021zst}. Unlike for black holes, however, the breakdown of the semi-classical description signals a fundamental incompatibility of de Sitter with quantum gravity~\cite{Dvali:2013eja,Dvali:2014gua,Dvali:2017eba,Dvali:2018fqu,Dvali:2018jhn,Dvali:2018ytn,Dvali:2020etd,Dvali:2021kxt}.}
 The latter effect originates from the enhanced memory capacity resulting from the large entropy~$S$ of a  black hole~\cite{Bekenstein:1973ur} -- in essence it is impossible to store the quantum information of an initial black hole of mass~$M$ in a smaller semi-classical black hole of mass~$M/2$ (with smaller entropy~$S/4$). Moreover, since the semi-classical Hawking evaporation cannot carry away any information, self-similarity must break down: \emph{Old black holes are fundamentally different from young black holes}~\cite{Dvali:2020wft,Alexandre:2024nuo}. 

A far-reaching question arises about the fate of a black hole beyond half evaporation. Two likely options are discussed in~\cite{Dvali:2020wft}; the black hole could either develop a classical instability, or experience a slowdown of evaporation. Analogies to other systems of enhanced memory storage capacity strongly point to the second option, meaning that memory burden backreacts against the decay and thereby stabilizes the black hole~\cite{Dvali:2020wft}. This makes observational searches for such long-lived small PBHs important for two reasons. Firstly, small PBHs become viable dark matter candidates if evaporation slows down since the lower bound on the PBH mass resulting from the requirement that they have survived until today drops below~$10^{14} \, \text{g}$~\cite{Dvali:2020wft}.  Secondly, detecting only a single small black hole of primordial origin in the present-day Universe would confirm that black hole evaporation slows down, thus greatly advancing our understanding of quantum gravity.

Whether or not they constitute a sizable fraction of dark matter, a slowdown of evaporation will significantly change the constraints on PBHs below~$10^{14} \, \text{g}$.\footnote
{Also, quasi-extremal black holes can evade the Hawking bound~\cite{deFreitasPacheco:2023hpb}.}
Following up on the first phenomenological considerations in~\cite{Dvali:2020wft}, a corresponding reevaluation of constraints was performed in~\cite{Alexandre:2024nuo,Thoss:2024hsr} and it was shown that the mass range between~$\sim 10^{6} \, \text{g}$ and~$\sim 10^{9} \, \text{g}$ is particularly interesting for dark matter. The upper bound is related to Big Bang Nucleosynthesis (BBN) -- for long-lived PBHs below~$\sim 10^{9} \, \text{g}$, it is certain that the slowdown of evaporation sets in before BBN, and so constraints that would otherwise follow from Hawking evaporation during BBN are alleviated completely. Regarding the lower end of the mass range, PBHs above~$\sim 10^{6} \, \text{g}$ can survive until today even if the slowdown of evaporation is comparatively `moderate', in a sense to be made precise below. Additionally, a slowdown of evaporation causes a model-independent spread of stabilized masses of initially degenerate PBHs~\cite{Dvali:2024hsb}.\footnote
{Both with~\cite{Haque:2024eyh,Barman:2024iht} and without~\cite{Anchordoqui:2024akj,Anchordoqui:2024dxu,Anchordoqui:2024jkn} memory burden, there can be further connections of long-lived small PBHs and dark matter.}

A leading mechanism to produce PBHs is from large cosmological perturbations that are seeded from quantum fluctuations produced during inflation. These perturbations collapse during radiation domination, producing PBHs with masses on the order of the horizon mass~\cite{Ivanov:1994pa,Hotchkiss:2011gz,Choudhury:2013woa} (see also~\cite{Garcia-Bellido:2017mdw,Motohashi:2017kbs,Germani:2017bcs,Ballesteros:2017fsr,Ballesteros:2020qam,Karam:2022nym,Bhatt:2022mmn}).\footnote
{Confinement provides another PBH production mechanism that does not require large inflationary fluctuations~\cite{Dvali:2021byy}.}
Although determining the precise abundance of PBHs remains a challenge due to an exponential sensitivity to the characteristics of inflationary perturbations~\cite{Franciolini:2018vbk,Ezquiaga:2018gbw,Cruces:2018cvq,Ezquiaga:2019ftu,Figueroa:2020jkf,Figueroa:2021zah,Hooshangi:2021ubn,Cai:2021zsp,Papanikolaou:2022chm,Animali:2022otk,Pi:2022ysn,Cruces:2022dom,Mishra:2023lhe,Kawaguchi:2023mgk,Hooshangi:2023kss,Stamou:2023vxu,Germani:2023ojx,Raatikainen:2023bzk,Choudhury:2024jlz,Mizuguchi:2024kbl,Escriva:2024aeo}, it is possible to constrain the amplitude of fluctuations for which PBHs can constitute all of dark matter within roughly one order of magnitude (see~\cite{Dandoy:2023jot}).\footnote
{Evidently, a large enhancement of perturbations is not only a blessing but can also be a curse since it inevitably limits the effectiveness of computational tools and endangers control over inflationary perturbation theory (see~\cite{Poisson:2023tja} and references therein).}
Large scalar fluctuations produce GWs at second order~\cite{Tomita:1967wkp} (see~\cite{Matarrese:1992rp,Matarrese:1993zf,Matarrese:1997ay,Ananda:2006af,Baumann:2007zm,Saito:2008jc,Assadullahi:2009jc,Bugaev:2009,Bugaev:2009a,Saito:2009,Bugaev:2010} for subsequent developments and~\cite{Domenech:2021ztg} for a review) which are possibly detectable by present and upcoming experiments. What is more, the dependence of second-order GWs on the properties of inflationary perturbations is not exponential and so the prediction of the GW signal associated with the production of PBHs is more robust than the abundance itself (as emphasized recently in~\cite{Pi:2024jwt}).  Therefore, our goal is to perform for the first time a simultaneous computation of inflationary observables and the GWs resulting from the inflationary production of dark matter in the form of long-lived PBHs below~$10^{14} \, \text{g}$.

Scalar-induced GWs associated with the production of small PBHs have already been investigated in~\cite{Franciolini:2023osw, Kohri:2024qpd}.\footnote{Other observational signatures of small PBHs can include high-energy neutrinos~\cite{Chianese:2024rsn} as well as observables related to their mergers, such as another GW signal with extremely high frequency~$\sim 10^{27} \, \text{Hz}$~\cite{Kohri:2024qpd} and ultrahigh-energy cosmic rays~\cite{Zantedeschi:2024ram}.} In these works, however, the spectrum of perturbations was not derived from an inflationary model: an analytic ansatz was assumed instead. Therefore, our work adds two crucial aspects to the results of~\cite{Franciolini:2023osw, Kohri:2024qpd}. First, we establish a connection of GW properties and inflationary observables. Although we shall rely on a particular model of inflation, there are good reasons to believe that the trends we observe, as described below, are generic. Second, the ans\"atze used in~\cite{Franciolini:2023osw, Kohri:2024qpd} feature free parameters that must be chosen ad hoc if the spectrum of perturbations is not derived from an inflationary model. Indeed, we shall show that fitting the functions used in~\cite{Franciolini:2023osw, Kohri:2024qpd} to the outcome of a concrete scenario of inflation yields other parameter values than the ones used in~\cite{Franciolini:2023osw, Kohri:2024qpd}. In turn, this difference has a significant influence on the GW spectrum and its future detectability.\footnote
{Also, without a connection to inflation, the effect of a slowdown of evaporation on GW signals has been studied in scenarios in which evaporation plays a significant role and PBHs do not assume the role of dark matter~\cite{Balaji:2024hpu,Barman:2024iht,Bhaumik:2024qzd,Barman:2024ufm}. In this case, an important GW signal can be induced by PBH number density fluctuations~\cite{Inomata:2020lmk,Papanikolaou:2020qtd,Domenech:2020ssp}, as also reviewed in~\cite{Domenech:2021ztg}. Moreover, we remark that GWs can also result from semi-classical evaporation \cite{Dong:2015yjs} and that the inflationary production of small evaporating PBHs was already studied in~\cite{Rasanen:2018fom,Basilakos:2023xof,Basilakos:2024diz}}.

	\begin{figure}
	\centering
	\includegraphics[width=1\textwidth]{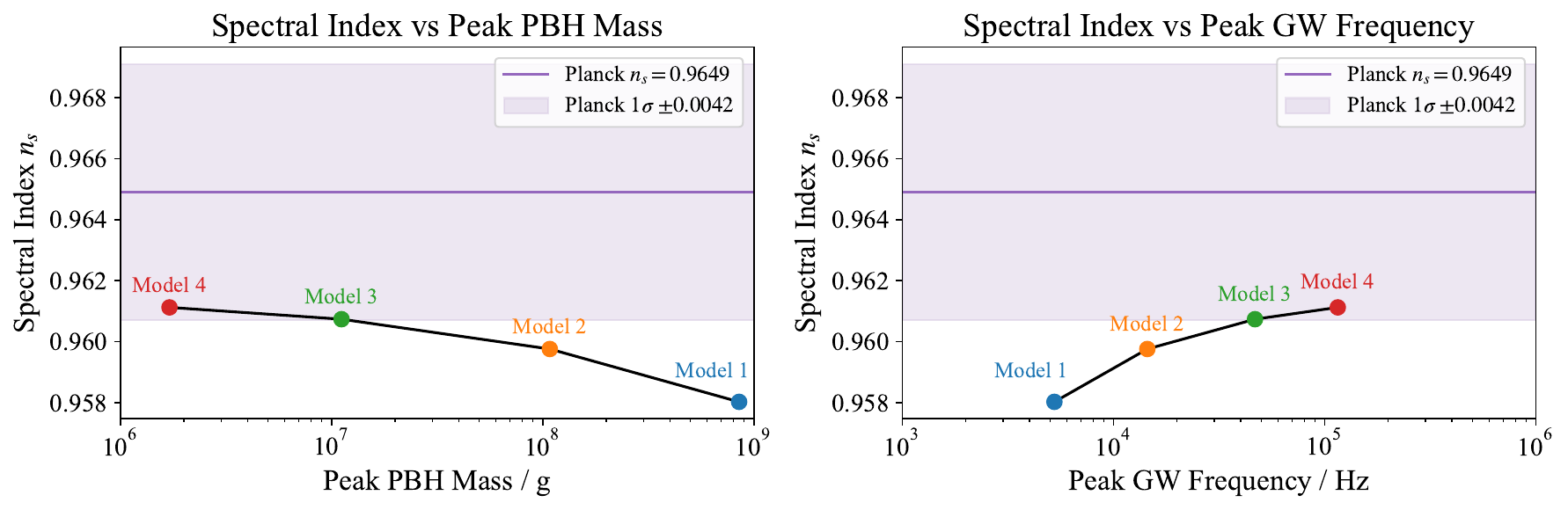}
	\caption{Spectral index~$n_s$ of the four models defined by~\cref{tab:model} plotted against the mass of the small PBHs which account for the entirety of dark matter (left panel), and the peak GW frequency (right panel). The observational constraints from Planck are also plotted~\cite{Planck:2018jri}.}
	\label{fig:spectral_index}
\end{figure}
For a concrete proof of concept, we will use a renormalizable polynomial inflaton potential with non-minimal coupling to gravity. This model, which among others is motivated by its similarity to Higgs inflation~\cite{Bezrukov:2007ep}, was introduced in~\cite{Ballesteros:2020qam} with the aim of producing PBHs above~$10^{14} \, \text{g}$. For this mass range, however, it was shown that a sizable abundance of PBHs cannot be produced while simultaneously satisfying constraints on inflation from observations of the cosmic microwave background (CMB)~\cite{Ballesteros:2020qam}.
We shall demonstrate that this tension disappears for PBHs of smaller masses. In~\cref{fig:spectral_index} we plot the spectral index of four models against the mass of the PBHs that account for the entirety of dark matter. Since the scalar spectral index still is at the lower bound of the currently allowed interval, future CMB observations have a potential to test the inflationary production of small PBHs.

	\begin{figure*}
	\center
	\includegraphics[width=\textwidth]{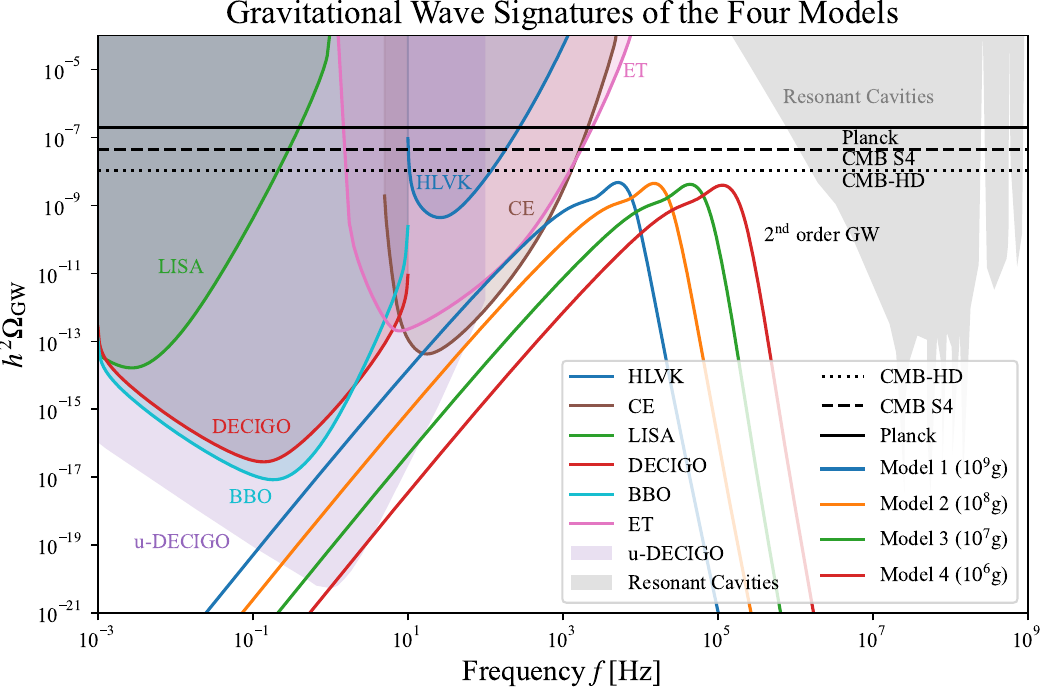}
	\caption{\label{fig:GWs} 
		Observational forecasts for the four new models proposed in this work. The plot shows the fraction of the energy density in GWs relative to the critical energy density of the Universe as a function of frequency, for models 1, 2, 3 and 4, as given in~\cref{tab:model}. For each model, small PBHs of mass~$10^9\,\mathrm{g}$,~$10^8\,\mathrm{g}$,~$10^7\,\mathrm{g}$ and~$10^6\,\mathrm{g}$ account for the totality of dark matter. We superimpose the power-law integrated sensitivity curves from~\cite{Schmitz:2020syl} for the prospective laser interferometers LISA~\cite{LISA:2017pwj, Baker:2019nia}, DECIGO~\cite{Kawamura:2006up, Yagi:2011wg, Isoyama:2018rjb}, BBO~\cite{Corbin:2005ny, Harry:2006fi}, CE~\cite{Reitze:2019iox}, ET~\cite{Punturo:2010zz}, and HLVK~\cite{KAGRA:2013rdx}. The ideal case of u-DECIGO is shaded in purple~\cite{Kuroyanagi:2014qza} (data taken from~\cite{Ringwald:2020vei}). Optimistic resonant cavity sensitivities are shaded in grey~\cite{Herman:2022fau} (data taken from~\cite{Drewes:2023oxg}). The black lines indicate the upper bounds on~$\Delta N_\mathrm{eff}$ of Planck~\cite{Planck:2018vyg} and its proposed successors, CMB-S4~\cite{CMB-S4:2016ple} and CMB-HD~\cite{CMB-HD:2022bsz}. Bounds are placed on the GW signature related to uncertainties in the PBH formation mechanism in~\cref{fig:GWBounds}, and the spectral index and tensor-to-scalar ratio are presented in~\cref{tab:model}. We see that there may be a possibility of detection of the signal's \emph{`shoulder'} for the u-DECIGO GW detector, and its peak comes close to the projected sensitivity of CMB-HD. For models producing PBHs with masses above~$10^8\,\mathrm{g}$, the GW signal can potentially be detectable with CE and ET.}
\end{figure*}
Regarding GW, we display our main result in~\cref{fig:GWs} for different models where PBHs with masses~$10^6\,\mathrm{g}$,~$10^7\,\mathrm{g}$,~$10^8\,\mathrm{g}$ or~$10^9\,\mathrm{g}$ constitute the totality of dark matter. We superimpose the expected power-law integrated sensitivity curves of future GW detectors~\cite{Corbin:2005ny,Harry:2006fi,Kawamura:2006up,Punturo:2010zz,Yagi:2011wg,KAGRA:2013rdx,Kuroyanagi:2014qza,LISA:2017pwj,Isoyama:2018rjb,Reitze:2019iox,Baker:2019nia, Herman:2022fau} on the stochastic GW background signal. Despite including optimistic forecasts, and neglecting important foregrounds effects, such as from black hole and neutron star binary mergers \cite{Franciolini:2023osw}, we find that the signal falls in a frequency region that appears to be largely unconstrained by current proposals of future GW detectors. Nevertheless, for models producing PBHs with masses above~$10^6\,\mathrm{g}$ there may be a possibility of detecting the signal's \emph{`shoulder'} for an ideal realisation of the u-DECIGO GW detector~\cite{Kuroyanagi:2014qza}. For models producing PBHs of masses above~$10^8\,\mathrm{g}$ there can be a potential for detection with Cosmic Explorer (CE)~\cite{Reitze:2019iox} and Einstein Telescope (ET)~\cite{Punturo:2010zz}. Finally, the peak of the signal comes close to the projected bound on~$\Delta N_{\text{eff}}$ resulting from a future CMB measurement by CMB-HD~\cite{CMB-HD:2022bsz}. Qualitatively our results agree with the findings of~\cite{Franciolini:2023osw, Kohri:2024qpd}, but the concrete bounds on the PBH mass for which GWs become detectable for a given forecast on experimental sensitivity are different, since the power spectrum that we derive from inflation differs from the ans\"atze used in~\cite{Franciolini:2023osw, Kohri:2024qpd}.\footnote
{Moreover, there is a question concerning the precise relationship between GW frequency and PBH mass (see detailed discussion in~\cref{app:pbh_production}).}
As another important outcome, our findings highlight the need to develop new experiments for probing GWs in the frequency range between~$\text{kHz}$ and~$\text{MHz}$ (see the Ultra-High-Frequency Gravitational Waves Initiative \cite{Aggarwal:2020olq}).

Our paper is structured as follows. In~\cref{sec:background}, we give more details about the slowdown of black hole evaporation and briefly review the inflationary model and findings of~\cite{Ballesteros:2020qam}.~\cref{sec:numerical} is devoted to our numerical approach for identifying model parameters that generate a significant abundance of PBHs. We present results for several such model parameterizations in~\cref{sec:results}, before concluding in~\cref{sec:conclusion}. In~\cref{app:background}, we review known results about the simultaneous production of PBHs and GWs from large inflationary perturbations, and we give details about the numerical procedure in~\cref{app:numerical}. 

	\textbf{Conventions:} We work in natural units~$\hbar\equiv c\equiv 1$, where~$M_P\equiv 1/\sqrt{8 \pi G}$ is the reduced Planck mass (with Newton's constant~$G$), and use the metric signature~$(-1,+1,+1,+1)$.

	\textbf{Comment:} Preliminary results associated with the present investigation appeared already in the Master's thesis~\cite{Gladwyn:2024} by BG.

\section{Background} \label{sec:background}

\subsection{Breakdown of Classical Description of Black Holes}
We will begin with a brief review of semi-classical particle production by a black hole and its potential breakdown. For a non-rotating, uncharged black hole of mass~$M$, and corresponding Schwarzschild radius~$r_g \equiv 2 G M$, we can form a single dimensionless quantity (keeping~$\hbar$ explicit for the present discussion)
\begin{equation} \label{S}
	S \equiv \frac{\pi r_g^2}{\hbar G} \;.
\end{equation}
Of course,~$S$ is the entropy of a black hole~\cite{Bekenstein:1973ur}, but the following argument can be understood without this identification. Even for small PBHs,~$S$ is huge (\eg~$S\gtrsim 10^{22}$ for~$M\gtrsim 10^6 \, \text{g}$). Semi-classical Hawking evaporation is fully determined by~$M$ and characterized by two quantities, namely the average energy of produced quanta~$E \sim \hbar r_g^{-1}$ and the rate of emission \cite{Hawking:1975vcx}
\begin{equation} \label{GammaSC}
	\Gamma_{\text{sc}} \sim  r_g^{-1} \;,
\end{equation}
where we did not display the greybody factor as well as the enhancement resulting from the total number of light degrees of freedom. If these properties remained unchanged throughout the entire evolution of a black hole, it would immediately follow that complete evaporation occurs after the emission of
\begin{equation} \label{N}
	N\sim \frac{M}{E} \sim S \;,
\end{equation}
Hawking quanta, corresponding to the lifetime
\begin{equation} \label{tSC}
	t_{\text{sc}} \sim \frac{N}{\Gamma_{\text{sc}}} \sim S r_g \;.
\end{equation}
Requiring~$t_{\text{sc}}$ to be larger than the age of the Universe immediately leads to the commonly-assumed Hawking bound that excludes PBHs below~$\sim 10^{14}\, \text{g}$ as viable dark matter candidates~\cite{Carr:2020xqk,Green:2020jor,Escriva:2022duf}.

However, the seemingly innocuous derivation of~\cref{tSC} is not self-consistent. The reason is that Hawking evaporation is computed semi-classically, \ie in the background of a \emph{fixed} black hole metric -- back-reaction is neglected~\cite{Page:1979}. The only way to make this approach exact is to employ the semi-classical limit~\cite{Dvali:2012en,Dvali:2012wq,Dvali:2013eja,Dvali:2015aja,Dvali:2017eba}
\begin{equation} \label{SCLimit}
	G\rightarrow 0 \;, \qquad M\rightarrow \infty \;, \qquad r_g=\text{const.}
\end{equation}
Obviously, a black hole of infinite mass does not experience back-reaction and so the metric will indeed not evolve. But at the same time, the semi-classical limit~\cref{SCLimit} implies that~$S\rightarrow \infty$, and so it is evident from~\cref{tSC} that~$t_{\text{sc}} \rightarrow \infty$. Not surprisingly, a black hole of infinite mass lives forever.

In order to make a statement about the lifetime of a black hole of finite mass, we need to extrapolate Hawking evaporation beyond the semi-classical limit. For a single emission, this leads to an error that scales with the back-reaction, \ie the relative change of the black hole mass~\cite{Dvali:2012rt,Dvali:2012en,Dvali:2012wq,Dvali:2013eja,Dvali:2015aja}:\footnote
{The same relative error can also be derived from the change of the semi-classical temperature~\cite{Dvali:2015aja}:~$\hbar\dot{T}/T^2 \sim 1/S$.}
\begin{equation}
	\frac{E}{M} \sim \frac{1}{S} \;.
\end{equation}
At first sight, this is good news. For cosmologically relevant black holes,~$1/S$ is tiny and so Hawking particle production is expected to be an extremely accurate approximation. However, the mass of a black hole only changes significantly after emitting~$\sim S$ quanta (see~\cref{N}). As a result, the total error resulting from the violation of the semi-classical limit is enhanced:
\begin{equation}
		S\, \frac{E}{M} \sim 1 \;.
\end{equation}
\emph{Once the mass of a black hole changes appreciably, the cumulative error becomes large and the semi-classical approximation can no longer be trusted ~\cite{Dvali:2012en,Dvali:2012wq,Dvali:2013eja,Dvali:2015ywa, Dvali:2015wca,Dvali:2015aja,Dvali:2016zqx,Dvali:2020wft}.} 

This conclusion fully resonates with earlier arguments showing that unitarity of black hole evolution requires a full deviation from Hawking evaporation at the latest after half of the mass is lost~\cite{Page:1993wv}. Evidently, it is possible that the semi-classical description breaks down much earlier, \eg already after~$\sim \sqrt{S}$ emission steps when the black hole has only lost a fraction~$\sim 1/\sqrt{S}$ of its initial mass~\cite{Dvali:2015wca,Michel:2023ydf}. In this case, a slowdown of evaporation would alleviate observational constraints even further as compared to a breakdown of semi-classicality after losing half of the mass~\cite{Thoss:2024hsr}. 

In summary, the fate of a black hole after half evaporation is an open question. In an attempt to answer it, quantum analogue models~\cite{Dvali:2012en,Flassig:2012re,Dvali:2013vxa,Dvali:2015ywa,Dvali:2015wca, Dvali:2016zqx,Dvali:2018tqi, Dvali:2018xpy,Dvali:2018xpy, Dvali:2020wft,Michel:2023ydf} were used that share important properties of black holes whilst allowing for easier theoretical study.  The fully non-perturbative numerical solution~\cite{Michel:2022kir} of a system that imitates black hole entropy yielded indications for a significant slowdown of evaporation corresponding to the reduced rate~\cite{Dvali:2020wft}
\begin{equation} \label{Gamma}
	\Gamma = \frac{1}{S^k} \Gamma_{\text{sc}} \;,
\end{equation}
where the precise value of~$k$ remains undetermined, roughly~$1\lesssim k \lesssim 3$. 

Key to the decelerating evaporation is \emph{memory burden}~\cite{Dvali:2018xpy,Dvali:2018xpy,Dvali:2020wft}, an effect that takes place in any system of enhanced memory capacity~\cite{Dvali:2021tez,Dvali:2023xfz,Dvali:2024hsb}. Its origin can be traced back to the sparsity of states that possess high microstate entropy. The requirement to maintain the high entropy at all points in time makes it impossible to depart from the state of enhanced memory capacity; the high entropy of the initial state counteracts any non-trivial evolution of the system.
This phenomenon can also be understood with the following semi-classical reasoning~\cite{Alexandre:2024nuo}: when the mass of a black hole decreases appreciably, also its entropy~\cref{S} shrinks significantly. Since Hawking evaporation is known to be featureless, unitarity requires a full deviation from semi-classical particle production.

For $k=1$, \ie a slowdown of evaporation by only a single power of entropy, PBHs with masses above~$\sim 10^{6}\, \text{g}$ survive until today~\cite{Alexandre:2024nuo,Thoss:2024hsr}.\footnote
{It is impossible to give a precise value for this lower bound without making further assumptions about the final stages of evaporation in the regime of slowdown. In~\cite{Thoss:2024hsr}, a lower bound around~$\sim 10^{7}\, \text{g}$ was used for~$k=1$.}
However, it follows from gamma ray constraints that only PBHs with masses above~$\sim 10^{9}\, \text{g}$ can account for the totality of dark matter~\cite{Thoss:2024hsr}, as is also discussed in \cite{Anchordoqui:2024dxu}. In the next-to-minimal case $k=2$, gamma ray constraints allow for masses above~$\sim 10^{6}\, \text{g}$~\cite{Thoss:2024hsr}, and therefore we shall focus on PBHs above this bound. Moreover, we will restrict ourselves to PBH masses below~$\sim 10^{9}\, \text{g}$, since in this case it is guaranteed that the slowdown of evaporation sets in before BBN and so constraints resulting from energy injection during BBN are essentially completely alleviated~\cite{Alexandre:2024nuo,Thoss:2024hsr}. It is important to emphasize that a slowdown of evaporation that sets in long before half of the mass is lost \cite{Michel:2023ydf} would further remove BBN constraints so that PBHs of masses above~$\sim 10^{9}\, \text{g}$ become viable dark matter candidates~\cite{Thoss:2024hsr}.

\subsection{Inflationary Prototype Model}
For our concrete model, we consider an inflaton field with a non-minimal quadratic coupling to the Ricci scalar in the metric formulation of GR. In the Jordan frame, the action takes the form~\cite{Birrell:1982ix,Hosotani:1985at,Madsen:1988ph}
\begin{equation}\label{Jordan}
S	= \int\mathrm{d}^4x\sqrt{-g}\biggl(\frac{M_P^2}{2}\Omega^2R - \frac{1}{2}g^{\mu\nu}\partial_\mu\phi\partial_\nu\phi-V(\phi)\biggr)\;,
\end{equation}
with~$\Omega^2\equiv1+\frac{\xi\phi^2}{M_P^2}$. Later we will choose~$V(\phi)$ to have a stationary inflection point a few e-folds before the end of inflation, so as to generate large inflationary perturbations. Keeping it general for now, we perform a Weyl transformation to the Einstein frame where coupling to gravity is minimal~\cite{Rigouzzo:2022yan, Carroll:2004st}. The metric transforms as~$\tensor{g}{_{\mu \nu}}\to\Omega^{-2}\tensor{g}{_{\mu \nu}}$ and the action of~\cref{Jordan} becomes
\begin{equation}\label{ActionTransformed}
	S= \int\mathrm{d}^4x\sqrt{-g}\biggl(\frac{M_P^2}{2}R - \frac{1}{2}K(\phi)g^{\mu\nu}\partial_\mu\phi\partial_\nu\phi-\Omega^{-4}V(\phi)\biggl) \;,
\end{equation}
where\footnote
{In other formulations of GR such as Palatini gravity, the second summand is absent,~$K(\phi)\equiv1/\Omega^2$ (see \eg~\cite{Rigouzzo:2022yan}). It would be interesting to study resulting inflationary scenarios, especially since Higgs inflation~\cite{Bezrukov:2007ep} is known to exhibit favorable properties in Palatini gravity~\cite{Bauer:2008zj,Bauer:2010jg,Rubio:2019ypq,Shaposhnikov:2020fdv,Dux:2022kuk,Poisson:2023tja}.}
\begin{equation}\label{KineticTransformed}
	K(\phi)\equiv\frac{1}{\Omega^2}+\frac{6[\Omega'(\phi)]^2}{\Omega^2} \;.
\end{equation}

As is usual, we canonically normalise the kinetic term by defining the field~$h$ such that~\cite{Poisson:2023tja}
\begin{equation}
	\frac{\mathrm{d}h}{\mathrm{d}\phi}\equiv\sqrt{K(\phi)}=\frac{\sqrt{1+\frac{\xi\phi^2(1+6\xi)}{M_P^2}}}{1+\frac{\xi\phi^2}{M_P^2}} \;.
\end{equation}
Using the boundary condition~$h(\phi=0)=0$, this can be readily integrated to~\cite{Ballesteros:2020qam}
\begin{equation}
	h\equiv\sqrt{\frac{1+6\xi}{\xi}}M_P\, \text{arcsinh}\left[\frac{\phi}{M_P}\sqrt{\xi(1+6\xi)}\right]
	-\sqrt{6}M_P\,\text{arctanh}{\left[\frac{\sqrt{6}\frac{\xi\phi}{M_P}}{\sqrt{1+\frac{\xi\phi^2(1+6\xi)}{M_P^2}}}\right]} \;.
\end{equation}
Now the theory in~\cref{ActionTransformed} is described by the action
\begin{equation} \label{canonicalInflaton}
	S= \frac{1}{2}\int\mathrm{d}^4x\sqrt{-g}\biggl(M_P^2R -\partial_\mu h\partial^\mu h-2U(h)\biggl) \;,
\end{equation}
 where the canonically normalised potential~$U(h)$ is expressed as
\begin{equation}
	U(h)\equiv \frac{V(\phi(h))}{\Omega^4(\phi(h))} \;.
\end{equation}
Standard inflationary analysis can be applied to~\cref{canonicalInflaton}, see~\cref{app:background}.

As adopted by~\cite{Ballesteros:2020qam}, we investigate the most general renormalizable potential for a real scalar~$\phi$ 
\begin{equation} \label{V}
    V(\phi)\equiv m^2 \phi^2+\mu\phi^3+\alpha \phi^4 \;,
\end{equation}
with an approximately stationary inflection point a few e-folds before inflation ends. We will choose the parameterization adopted in~\cite{Ballesteros:2020qam,Ballesteros:2017fsr} so that the inflection point occurs explicitly at~$\phi_0$
\begin{equation}\label{param}
    V(\phi) = \frac{\lambda \phi^4}{4!}\Bigg[3 + \xi^2 \frac{\phi_0^4}{M_P^4} -8(1+c_3)\frac{\phi_0}{\phi}+2(1+c_2)\Big(3 + \xi\frac{\phi_0^2}{M_P^2}\Big)\frac{\phi_0^2}{\phi^2}\Bigg]\;.
\end{equation}
The authors of~\cite{Ballesteros:2020qam} identified a 3$\sigma$ tension between the measured and expected spectral index for models which (i) produced PBHs with a mass just above the semi-classical Hawking evaporation limit and (ii) could account for the entirety of dark matter.\footnote
{This tension can also be alleviated in a non-minimal model, in which non-renormalizable operators are added to the potential~\cref{V}~\cite{Ballesteros:2020qam}, and in multi-field generalizations of the model \eqref{Jordan}~\cite{Geller:2022nkr,Qin:2023lgo}.}
In the present paper, we shall look for potentials that produce a PBHs of mass~$10^6-10^9\,\mathrm{g}$ that account for the entirety of dark matter.

\section{Numerical Approach}\label{sec:numerical}
\begin{figure}[h]
    \center
    \includegraphics[height=0.85\textheight]{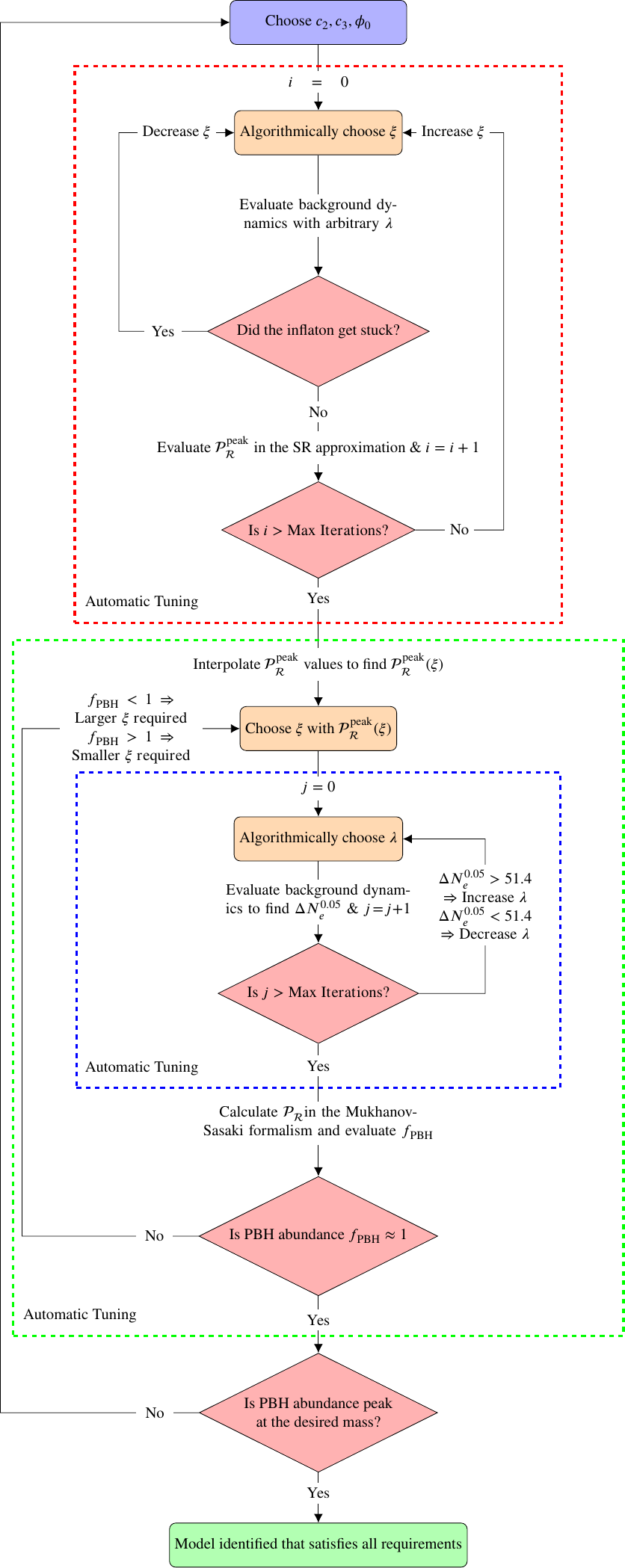}
    \caption{\label{fig:flowchart} 
	An outline of the approach used to identify potentials that satisfy all the requirements set out in~\cref{sec:requirements}. The dashed boxes represent automatic tuning procedures performed by a computer.
	}
\end{figure}

We adjust the five model parameters to meet specific requirements on the PBH mass and abundance. We must ensure compatibility with the CMB normalization and that the model solves the horizon and flatness problems~\cite{Guth:1980zm, Baumann:2022mni}. An outline of the tuning procedure is represented in~\cref{fig:flowchart}, and key points are highlighted in the following sections. Details of the methods used to control numerical instabilities that occurred during the tuning procedure are given in~\cref{app:instabilities}. 

\subsection{Model Requirements}\label{sec:requirements}
In our analysis, we seek potentials that satisfy the following requirements:
\begin{enumerate}
    \item \emph{Compatibility with CMB constraints on the comoving curvature perturbation spectrum.}\vspace{0.1cm}\\
    Specifically, we want compatibility with the constraints imposed by the Planck and BICEP CMB anisotropy measurements~\cite{Planck:2018jri,BICEP:2021xfz} on the spectral index~$n_s$, tensor-to-scalar ratio~$r$, and scalar amplitude~$A_s$ at the pivot scale~$k_*=0.05\mathrm{Mpc}^{-1}$ (see~\cref{app:perturbations} for definitions). To make contact with observational scales, we rescale the wavenumbers~$k$ so that we have~$k=k_*$ where the magnitude of the comoving curvature perturbation spectrum~$\mathcal{P}_\mathcal{R}(k)$ equals the Planck CMB measurement of the scalar amplitude. This immediately guarantees compatibility of~$A_s$ at the pivot scale with the Planck measurements. In contrast, we consider~$n_s$ and~$r$ as outputs of the model, \ie we do not use observational information about~$n_s$ and~$r$ to adjust model parameters.
    \item \emph{Solve the horizon and flatness problems}\vspace{0.1cm}\\
    The modes measured in the CMB were produced approximately ~$\Delta N^{0.05}_e \simeq 50 - 60$ e-folds before the end of inflation~\cite{Baumann:2022mni}. 
    The precise value of~$\Delta N^{0.05}_e$ depends on the energy scale of inflation (which has not been measured yet, only constrained with the tensor-to-scalar ratio~\cite{1711.07344}) and the dynamics of reheating~\cite{1405.5538}.~$\Delta N^{0.05}_e=51.4$ was chosen to check for consistency with~\cite{Ballesteros:2020qam} before obtaining original results, and is subsequently adopted throughout. We use the rescaling degree of freedom in the potential,~$\lambda$, to ensure this condition is met. 
    \item \emph{A peak in the comoving curvature perturbation spectrum~$\mathcal{P}_\mathcal{R}(k)$ that generates a mass fraction of PBH as dark matter of~$f_\mathrm{PBH}\approx1$ in the mass range of interest~$10^6-10^9\,\mathrm{g}$.}\vspace{0.1cm}\\
    To produce~$f_\mathrm{PBH} \sim 1$ for PBHs of~$10^6 - 10^9\,\mathrm{g}$ we require a peak in~$\mathcal{P}_\mathcal{R}(k)$ of~$\mathcal{P}^\mathrm{peak}_\mathcal{R}(k)\simeq10^{-2}$ between the wavenumbers~$10^{18}-10^{20}\,\mathrm{Mpc}^{-1}$. The wavenumber at the peak is controlled by~$c_2$,~$c_3$ and~$\phi_0$. The amplitude requirement is satisfied by tuning~$\xi$.
\end{enumerate}
We find model parameters that satisfy these requirements with a mixture of manual and procedural tuning. The outputs~$n_s$ and~$r$ are used to test the model against observations.

\subsection{PBH Mass and Abundance Tuning}
To identify a model that satisfies the PBH abundance and mass criteria of requirement 3, we adjust the parameters~$c_2$,~$c_3$ and~$\phi_0$. Critically,~$\phi_0$ controls the position of the inflection point for the limit~$c_2=0$,~$c_3=0$. Therefore, reducing~$\phi_0$ brings the stationary point closer to~$\phi=0$, which delays the ultra-slow-roll (USR) phase to later in inflation. This increases the wavenumber of the peak in~$\mathcal{P}_\mathcal{R}(k)$, and correspondingly makes the PBH smaller. Starting with the values used in~\cite{Ballesteros:2020qam} we increased~$c_2$ and~$c_3$ to combat the apparent increase in difficulty tuning~$\xi$, which is associated with decreasing~$\phi_0$.

For a given set of parameters~$c_2$,~$c_3$ and~$\phi_0$, an automated procedure, represented in the red dashed box in~\cref{fig:flowchart}, is used to identify values of~$\xi$ that boost~$\mathcal{P}^\mathrm{peak}_\mathcal{R}$ sufficiently in order to satisfy~$f_\mathrm{PBH}\simeq 1$. Increasing~$\xi$ acts to increase the depth of the local minimum in the potential, slowing the inflaton velocity and boosting the power spectrum. However, if~$\xi$ becomes too large then the inflaton becomes stuck at the local minimum and inflation never ends. To boost~$\mathcal{P}^\mathrm{peak}_\mathcal{R}$, we must fine-tune the parameter~$\xi$ to remain just below the critical value at which the inflaton becomes stuck~$\xi^\mathrm{critical}$, typically to within~$\frac{\xi^\mathrm{critical}-\xi}{\xi^\mathrm{critical}}\lesssim10^{-6}$.

The automatic procedure trials different values of~$\xi$, increasing or decreasing~$\xi$ depending on whether the inflaton becomes stuck. The equations governing the computation of the trials are given in~\cref{app:background} and details on the implementation of our approach are given in~\cref{app:tuning_xi}. During the tuning procedure, an arbitrary first guess for~$\lambda$ of~$1\times10^{-9}$ is used. It is useful to evaluate the power spectrum in the SR approximation for each successful calculation of the background dynamics (where the inflaton does not become stuck) and identify the peak value~$\mathcal{P}^\mathrm{peak}_\mathcal{R}(\xi)$. We can then interpolate~$\mathcal{P}^\mathrm{peak}_\mathcal{R}(\xi)$ to find~$\xi$ for a desired~$\mathcal{P}^\mathrm{peak}_\mathcal{R}$.

For a given~$\xi$,~$\lambda$ is tuned so as to achieve the desired number of e-folds~$\Delta N^{0.05}_e=51.4$ between the pivot scale exiting the horizon and the end of inflation, thus satisfying requirement 2. This process is represented in the blue dashed box in~\cref{fig:flowchart}. Once~$\lambda$ has been determined, the PBH abundance~$f_\mathrm{PBH}$ can be evaluated. We employ the simple Press-Schechter formalism described in~\cref{app:pbh_production} (as in~\cite{Ballesteros:2020qam}). More accurate approaches have been developed, but significant uncertainties remain even in with these more sophisticated methods~\cite{Franciolini:2018vbk,Ezquiaga:2018gbw,Cruces:2018cvq,Ezquiaga:2019ftu,Figueroa:2020jkf,Figueroa:2021zah,Hooshangi:2021ubn,Cai:2021zsp,Papanikolaou:2022chm,Animali:2022otk,Pi:2022ysn,Cruces:2022dom,Dandoy:2023jot,Mishra:2023lhe,Kawaguchi:2023mgk,Hooshangi:2023kss,Stamou:2023vxu,Germani:2023ojx,Raatikainen:2023bzk,Choudhury:2024jlz,Mizuguchi:2024kbl,Escriva:2024aeo}. The Press-Schechter formalism makes it much easier to account for the unavoidable uncertainties in the calculation of PBH-abundances (see~\cref{sec:bounds}). As a final step,~$\xi$ is then tuned until we have identified~$\xi$ and~$\lambda$ that are consistent with the CMB normalisation and produces a peak in~$\mathcal{P}_\mathcal{R}(k)$ such that~$f_\mathrm{PBH}\approx1$. This is implemented in an automatic tuning procedure shown in the green dashed box in~\cref{fig:flowchart}.

\subsection{Gravitational Wave Bounds}\label{sec:bounds}
Large theoretical uncertainties in the calculation of PBH-adundances~\cite{Franciolini:2018vbk,Ezquiaga:2018gbw,Cruces:2018cvq,Ezquiaga:2019ftu,Figueroa:2020jkf,Figueroa:2021zah,Hooshangi:2021ubn,Cai:2021zsp,Papanikolaou:2022chm,Animali:2022otk,Pi:2022ysn,Cruces:2022dom,Dandoy:2023jot,Mishra:2023lhe,Kawaguchi:2023mgk,Hooshangi:2023kss,Stamou:2023vxu,Germani:2023ojx,Raatikainen:2023bzk,Choudhury:2024jlz,Mizuguchi:2024kbl,Escriva:2024aeo} make it impossible to derive a unique prediction for the GW signals that accompany the production of a given abundance of small PBHs. Going beyond previous investigations (see~\cite{Franciolini:2023osw,Kohri:2024qpd}), we shall therefore place upper and lower bounds on the stochastic GW background of models that may allow for producing a PBH abundance that fully accounts for dark matter.\footnote
{The simultaneous production of scalar-induced GWs and small PBHs affected by memory burden was also recently studied in \cite{Jiang:2024aju}, with the finding that some mass ranges could already be ruled out by present GW data. In that work, however, uncertainties in the calculation of PBH abundances were not taken into account.}

In order to derive these bounds, we shall use the following uncertainties that are already evident in the Press-Schechter formalism. Firstly, the mass fraction of PBHs~$f_\mathrm{PBH}$ depends exponentially on the critical threshold~$\delta_c$. Secondly,~$f_\mathrm{PBH}$ has an exponential dependence on the variance, which depends strongly on the smoothing window function~$W(x)$, for which there is no unique definition. The uncertainty in the choices of~$\delta_c$ and~$W(x)$ results in an uncertainty in the size of the required peak in~$\mathcal{P}_\mathcal{R}$ so that we have~$f_\mathrm{PBH}\approx1$. This uncertainty in the GW signature is approximately quadratic in the uncertainty of~$\mathcal{P}_\mathcal{R}$ (see~\cref{app:gwcalc}). Therefore, we place bounds on the GW spectrum by retuning the comoving curvature perturbation~$\mathcal{P}_\mathcal{R}$ (by changing~$\xi$, which controls the size of the peak) so that we have~$f_\mathrm{PBH}=1$ for different choices of the threshold~$\delta_c$ and window function~$W(x)$.

A lower bound is found by taking~$\delta_c=0.4$ and using a top-hat window function, since this estimate generates the most PBHs. Therefore, the peak in~$\mathcal{P}_\mathcal{R}(k)$ is the smallest necessary to achieve~$f_\mathrm{PBH}=1$, and so the GW signal is also the smallest. An upper bound is found by taking~$\delta_c=0.6$ and using a Gaussian window function, since this produces the smallest number of PBHs meaning that we must enhance~$\mathcal{P}_\mathcal{R}(k)$ for~$f_\mathrm{PBH}=1$. The peak in~$\mathcal{P}_\mathcal{R}(k)$ is therefore the largest, and so the GW signal is also the largest.

\section{Results} \label{sec:results}
Using the numerical approach outlined in~\cref{sec:numerical}, we identify several potentials that satisfy our requirements in~\cref{sec:requirements}. We present four models that produce PBH abundance peaks that span the mass range of interest in~\cref{tab:model}, along with their corresponding spectral index~$n_s$ and tensor-to-scalar ratio~$r$.\footnote
{Already in flat space, a necessary condition for the theory \eqref{V} to be weakly coupled is that the strengths of the 3- and 4-point interactions fulfill~$\mu < m$ and~$\alpha<1$, respectively. Translating our results shown in~\cref{tab:model} from the parametrization \eqref{param} back to~\cref{V}, we have checked that all four models obey these two weak-coupling conditions.}

\begin{table}[h!]
\centering
\renewcommand{\arraystretch}{1.3} 
\begin{tabular}{c|c|c|c|c|c|c|c|c}
    \hline
    \hline
    \textbf{Model} & \textbf{PBH mass / g} & \textbf{\( c_2 \)} & \textbf{\( c_3 \)} & \textbf{\( \phi_0 \)} & \textbf{\( \xi \)} & \textbf{\( \lambda \)} & \textbf{\( n_s \)} & \textbf{\( r \)} \\ \hline
    1 & \(\sim10^9\) & 0.03 & 0.075 & 0.97 &~$\sim 0.10$ &~$\sim 6.6\times10^{-11}$ & 0.9580 & 0.021 \\ \hline
    2 & \(\sim10^8\) & 0.03 & 0.075 & 0.82 &~$\sim 0.24$ &~$\sim 2.6\times10^{-10}$ & 0.9598 & 0.016 \\ \hline
    3 & \(\sim10^7\) & 0.03 & 0.075 & 0.65 &~$\sim 0.53$ &~$\sim 1.1\times10^{-9}$ & 0.9608 & 0.014 \\ \hline
    4 & \(\sim10^6\) & 0.05 & 0.075 & 0.05 &~$\sim 130$ &~$\sim 5.3\times10^{-5}$ & 0.9611 & 0.012 \\ \hline
\hline
\end{tabular}
	\caption{Table of parameters identifying the four models and their corresponding spectral index \( n_s \) and tensor-to-scalar ratio \( r \). As outlined in~\cref{sec:numerical}, the parameters~$\xi$ and~$\lambda$ have been tuned, and approximate values are given here. See supplemental materials at~\cite{Supplement} for full numerical details.}
\label{tab:model}
\end{table}

\begin{figure}
  \centering
  \includegraphics[width=1\linewidth]{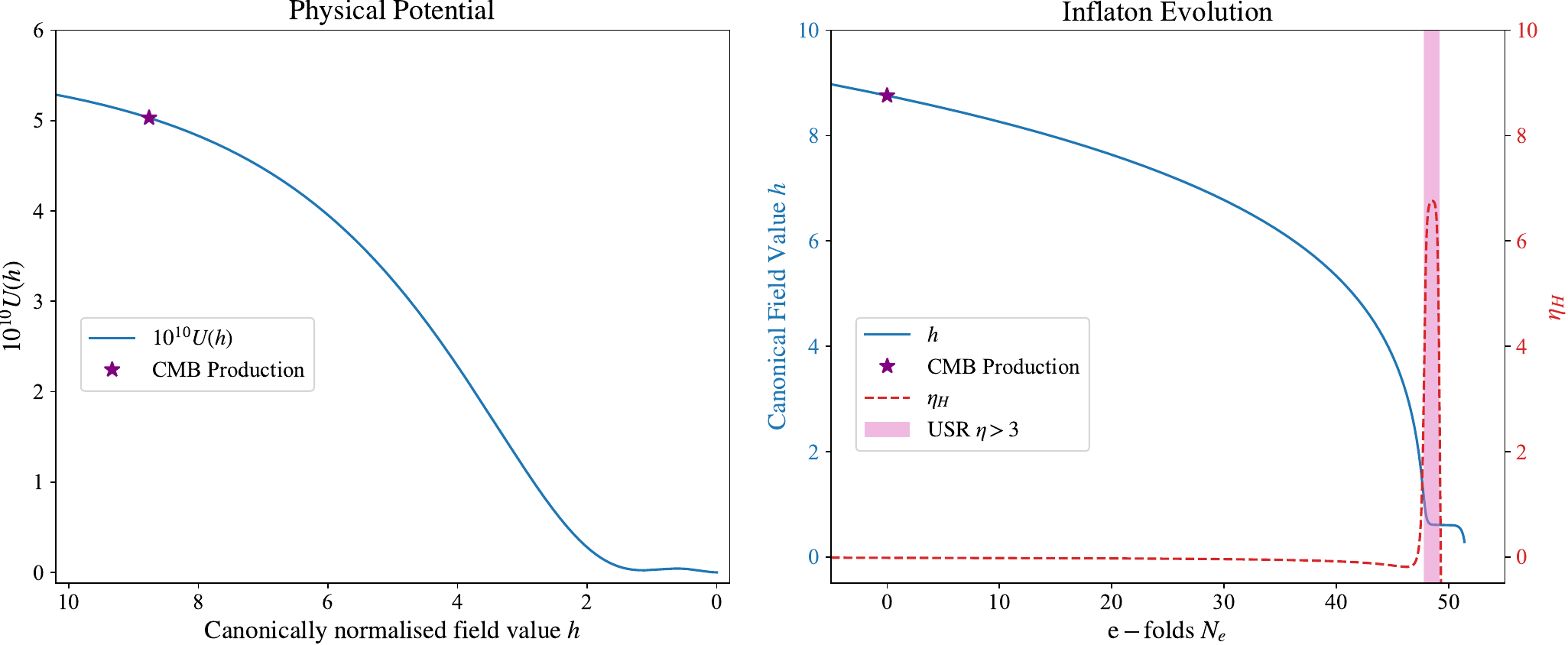}
	\caption{\label{fig:inflatonandpotential} 
	The left panel shows the physical potential \( U(h) \) as a function of the canonically normalized field value \( h \) for model~2 of~\cref{tab:model}. The right panel shows the evolution of the canonically normalized field value \( h \) until the end of inflation for the same model. Parameters \( \xi \) and \( \lambda \) are tuned to enhance the comoving power spectrum and achieve the correct CMB normalization. The CMB production at the pivot scale \( k_*=0.05\, \mathrm{Mpc}^{-1} \) is marked with a purple star. The slow-roll parameter \( \eta_H \) is shown, and the ultra-slow-roll (USR) phase where \( \eta_H > 3 \) is highlighted in pink. A purple star indicates CMB production at the pivot scale in both plots.
	}
\end{figure}

We will first present the background evolution and cosmological perturbations of model~2 and will later show the PBH abundances and gravitational wave signatures of all four. The physical potential~$U(h)$ is plotted as a function of the canonically normalised field~$h$ in the left panel of~\cref{fig:inflatonandpotential}. We see a shallow minimum which slows the inflaton field velocity and boosts the peak in the comoving curvature perturbation spectrum high enough to produce a significant abundance of PBHs. In the right panel of~\cref{fig:inflatonandpotential}, we plot the dynamics of~$h$ calculated with~\cref{eom} and SR initial conditions. We plot~$\eta_H$ to indicate the USR phase ($\eta_H>3$) that is highlighted in pink. The CMB production at the pivot scale is indicated with a purple star.

\subsection{Background Evolution and Perturbations}
\begin{figure}
  \center
  \includegraphics[width=\linewidth]{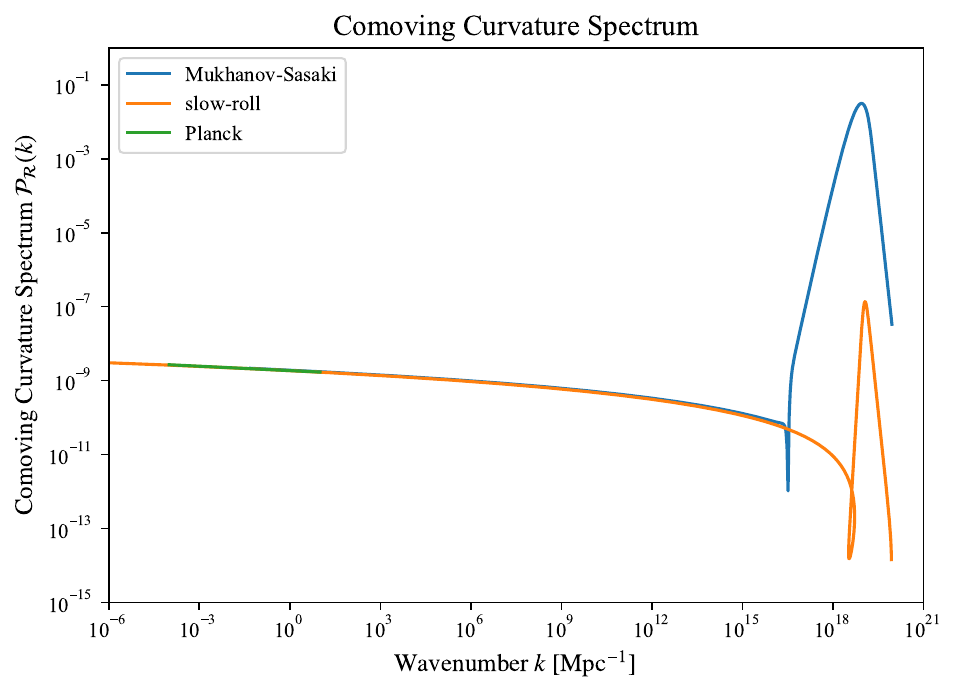}
	\caption{\label{fig:powerspectrum} 
	Power spectrum of comoving curvature perturbations as a function of wavenumber~$k$ calculated with the Mukhanov-Sasaki formalism (blue) and in the SR approximation (orange) for the model~2 shown in~\cref{tab:model}. The measured spectrum from Planck is plotted to first order (green). The spectral index is~$n_s=0.9598$ and the tensor-to-scalar ratio is~$r=0.016$, both of which are in good agreement with measurements from Planck~\cite{Planck:2018jri} and BICEP~\cite{BICEP:2021xfz}.
	}
\end{figure}

The comoving curvature spectrum is calculated in the SR approximation and separately with the Mukhanov-Sasaki formalism (see~\cref{app:perturbations} for equations). The results are plotted in~\cref{fig:powerspectrum}. We observe a sudden dip followed by a large enhancement during the USR phase. The five orders of magnitude difference between the Mukhanov-Sasaki calculation and the SR approximation are especially significant due to the exponential sensitivity of~$f_\mathrm{PBH}$ on~$\mathcal{P}_\mathcal{R}(k)$.

To compare our model to the CMB observables, we calculate the scalar index~$n_s$, and tensor-to-scalar ratio~$r$ (recall that~$A_s$ has been fixed to the measured value already, so it is not an output of our model). Since at CMB scales the power spectrum is well-described by the SR approximation, we calculate these quantities with the slow-roll formulae in~\cref{spectralindex}, and~\cref{tensorscalar}, respectively. The results are shown in~\cref{tab:model} and are plotted as a function of PBH mass in~\cref{fig:spectral_index}. We observe that the spectral index is still at the lower bound of the Planck measurements~\cite{Planck:2018jri}, and small PBHs (and therefore higher frequency GW signals) allow for better agreement with Planck~\cite{Planck:2018jri} at the~$1\sigma$ level. The tensor-to-scalar ratio is also consistent with the BICEP measurement~$r\lesssim0.04$~\cite{BICEP:2021xfz}.

\subsection{Gravitational Wave Spectrum}
\begin{figure*}[p]
    \centering
    \begin{tabular}{cc}
        \includegraphics[width=0.5\textwidth]{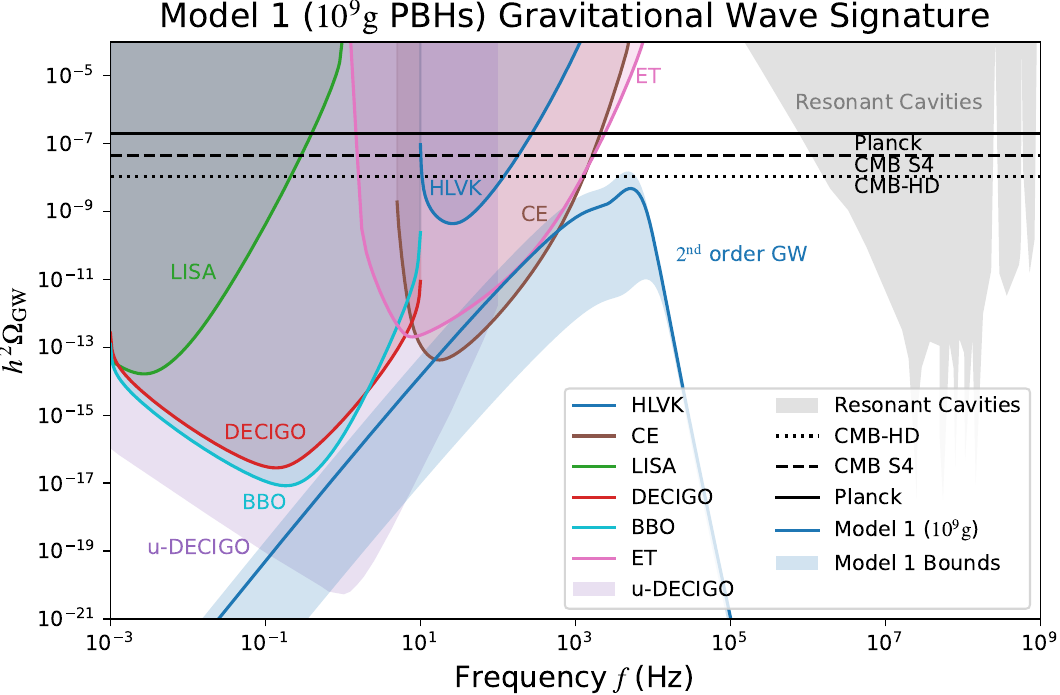} & 
        \includegraphics[width=0.5\textwidth]{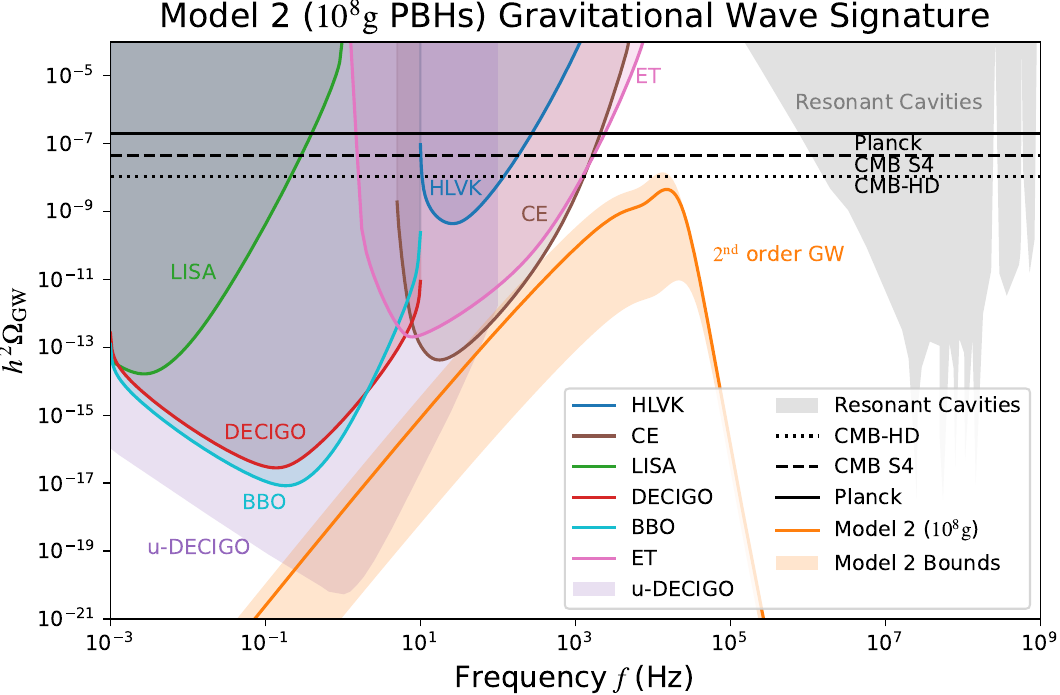} \\
        \includegraphics[width=0.5\textwidth]{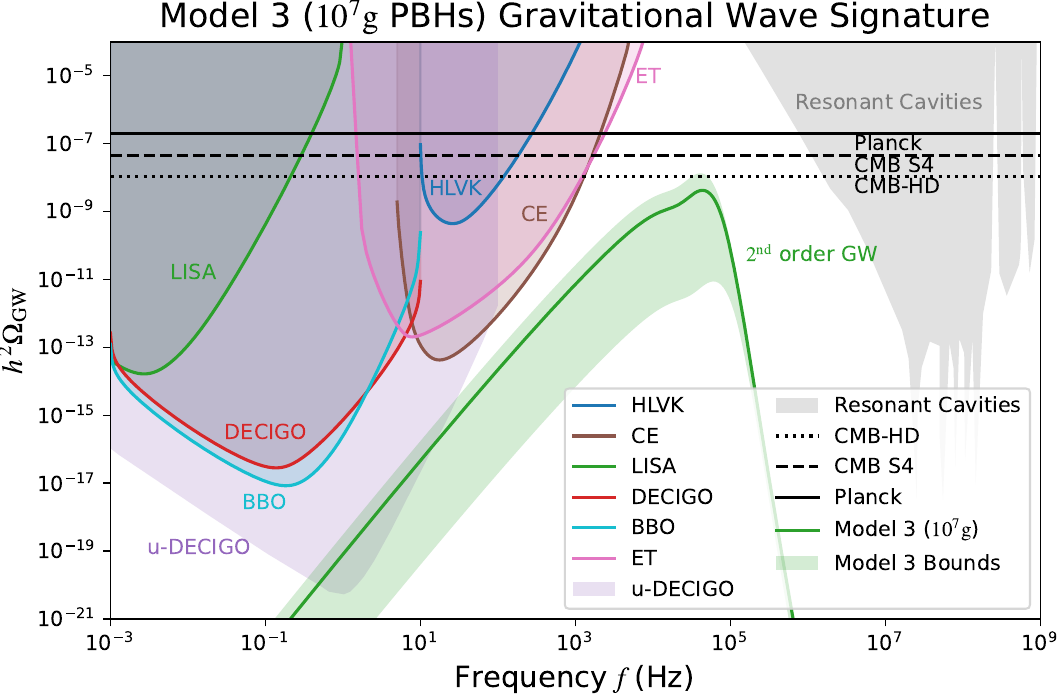} & 
        \includegraphics[width=0.5\textwidth]{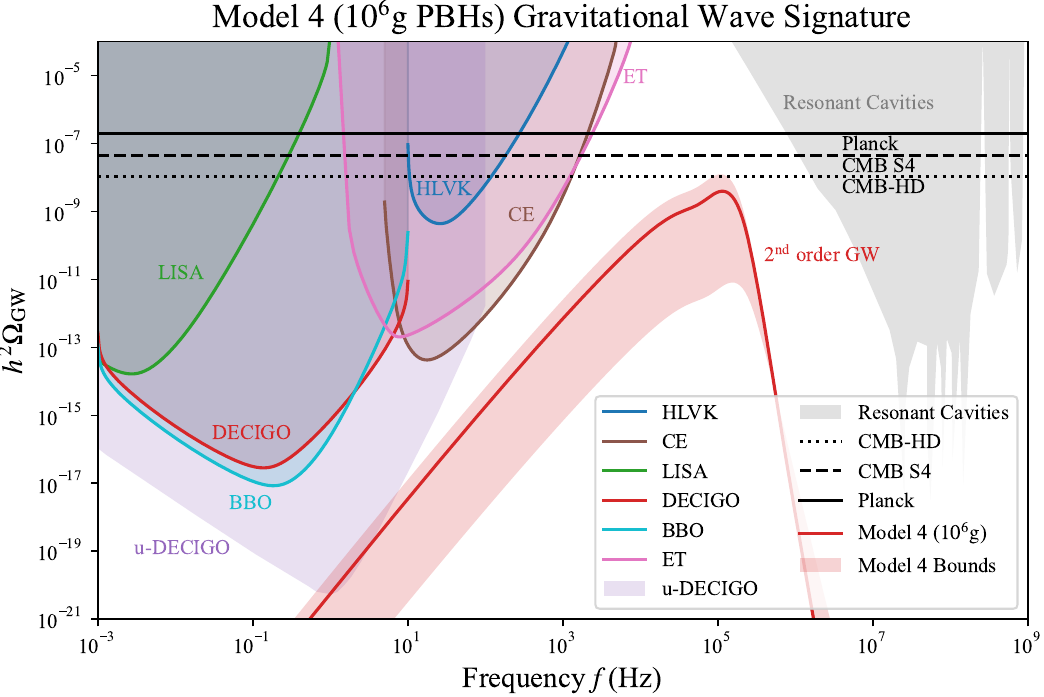} \\
    \end{tabular}

    \caption{\label{fig:GWBounds} Observational forecasts for the four new models proposed in this report. The plots show the fraction of the energy density in gravitational waves relative to the critical energy density of the Universe as a function of frequency for models 1, 2, 3, and 4 given in \cref{tab:model}. Bounds are placed upon the GW signature as described in~\cref{sec:bounds}. For each model, small PBHs with mass~$10^9\, \mathrm{g}, 10^8\, \mathrm{g}, 10^7\, \mathrm{g}$, and~$10^6\, \mathrm{g}$ account for the totality of dark matter. We superimpose the power-law integrated sensitivity curves from~\cite{Schmitz:2020syl} for the prospective laser interferometers LISA~\cite{LISA:2017pwj, Baker:2019nia}, DECIGO~\cite{Kawamura:2006up, Yagi:2011wg, Isoyama:2018rjb}, BBO~\cite{Corbin:2005ny, Harry:2006fi}, CE~\cite{Reitze:2019iox}, ET~\cite{Punturo:2010zz}, and HLVK~\cite{KAGRA:2013rdx}. The ideal case of u-DECIGO is shaded in purple~\cite{Kuroyanagi:2014qza} (data taken from~\cite{Ringwald:2020vei}). Optimistic resonant cavity sensitivities are shaded in grey~\cite{Herman:2022fau} (data taken from~\cite{Drewes:2023oxg}). The black lines indicate the upper bounds on~$\Delta N_\mathrm{eff}$ of Planck~\cite{Planck:2018vyg} and its proposed successors, CMB-S4~\cite{CMB-S4:2016ple} and CMB-HD~\cite{CMB-HD:2022bsz}. All four models may have a potential for detection of the signal's \emph{`shoulder'} by the u-DECIGO GW detector, and depending on the uncertainty in the GW signal its peak can be reachable by CMB-HD. Moreover, for models producing PBHs with masses above~$10^8\,\mathrm{g}$ there may be a possibility for detection with CE and ET.}
\end{figure*}

We compute the stochastic GW background using the calculation described in~\cref{app:gwcalc}. The GW energy density in units of the critical energy density and as a function of frequency is plotted in~\cref{fig:GWs} and bounds in~\cref{fig:GWBounds}. We superimpose the power-law integrated sensitivity curves from~\cite{Schmitz:2020syl} for the prospective laser interferometers LISA~\cite{LISA:2017pwj, Baker:2019nia}, DECIGO~\cite{Kawamura:2006up, Yagi:2011wg, Isoyama:2018rjb}, BBO~\cite{Corbin:2005ny, Harry:2006fi}, CE~\cite{Reitze:2019iox}, ET~\cite{Punturo:2010zz}, and HLVK~\cite{KAGRA:2013rdx}. The ideal case of u-DECIGO is shaded in purple~\cite{Kuroyanagi:2014qza} (data taken from~\cite{Ringwald:2020vei}). Power-law integrated sensitivity curves include the additional sensitivity that comes from the broadband nature of the stochastic GW signal via the integration over frequency~\cite{Thrane:2013oya}. We do not discuss effects of astrophysical foregrounds, \eg from black hole and neutron star binary mergers, which may however play an important role for some of the experiments shown in our \cref{fig:GWs,fig:GWBounds} \cite{Franciolini:2023osw}. The region shaded in grey shows the sensitivity estimate of a recently proposed resonant cavity for high-frequency GWs~\cite{Herman:2022fau}.\footnote
{Resonant cavity~\cite{Herman:2022fau} data taken from~\cite{Drewes:2023oxg} and converted from characteristic strain~$h_c$ to dimensionless energy density~$h^2\Omega_\mathrm{GW}$~\cite{Moore:2014lga}.}
 It remains to be clarified, however, if this sensitivity forecast for resonant cavities may be too optimistic (see~\cite{Navarro:2023eii}).
Several proposals for experiments operating in the correct frequency range including LSD, DMR and BAW but have projected sensitivities which are dramatically too low, and are therefore not plotted~\cite{Aggarwal:2020olq, Franciolini:2022htd,Arvanitaki:2012cn,Domcke:2022rgu,Goryachev:2014yra}. The constraints on the stochastic GW background from~$\Delta N_\mathrm{eff}$ measurements by Planck~\cite{Planck:2018jri} are plotted along with its proposed successors, CMB-S4~\cite{CMB-S4:2016ple} and CMB-HD~\cite{CMB-HD:2022bsz}.

The bounds placed on the model span three orders of magnitude, which is significantly less than the uncertainty associated with the PBH production mechanism but still affects the observational forecasts. While in an optimistic case all models producing PBHs with masses above~$10^6\,\mathrm{g}$ may lead to a potential of detecting the signal's \emph{`shoulder'} for an ideal realisation of u-DECIGO GW detector~\cite{Kuroyanagi:2014qza}, this is only possible for masses above~$10^9\,\mathrm{g}$ in the most pessimistic forecasts. Likewise, only in the most optimistic case all models are detectable by CMB-HD~\cite{CMB-HD:2022bsz}. For models producing PBHs with masses above~$10^8\,\mathrm{g}$, a potential for detection with CE~\cite{Reitze:2019iox} and ET~\cite{Punturo:2010zz} can arise, but again this is sensitive to the uncertainty in the GW amplitude. In any case, the peak of the signal falls in a frequency domain where there is no projected sensitivity in current proposals for future GW detectors Our results highlight the need to develop experimental approaches in the frequency domain around~$10^5\,\mathrm{Hz}$ -- see also the related Ultra-High-Frequency Gravitational Waves Initiative \cite{Aggarwal:2020olq}.

\clearpage
\clearpage

\subsection{PBH Abundances}
\begin{figure}[!t]
	\center
	\includegraphics[width=1\linewidth]{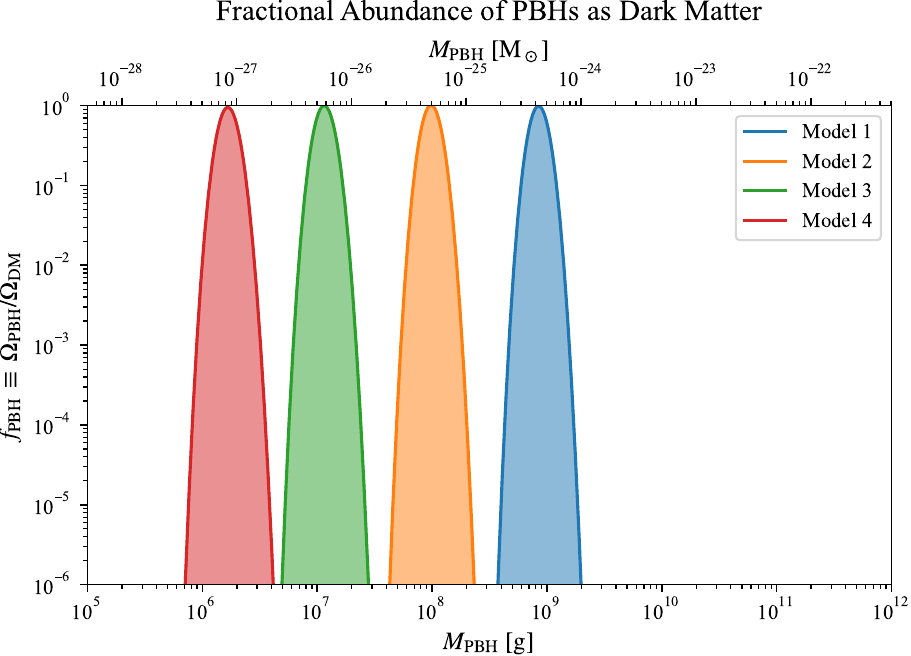}
	\caption{\label{fig:pbh_abundance_plot} 
		Fractional abundance of PBHs with respect to dark matter abundance as a function of PBH mass for the four models defined in~\cref{tab:model}. The calculation follows the Press–Schechter formalism reviewed in~\cref{app:pbh_production} for a choice of~$\delta_c=0.45$ and a Gaussian window function.
	}
\end{figure}

Finally, we perform a calculation of PBH abundances as a function of PBH mass using the Press–Schechter formalism reviewed in~\cref{app:pbh_production}. We use~$\delta_c=0.45$, and a Gaussian window function as our best estimate. The result is shown in~\cref{fig:pbh_abundance_plot}, where we remind the reader that we have tuned the parameter~$\xi$ to that ensure that we produce an~$O(1)$ PBH abundance. It is important to reiterate that the abundances computed using the Press–Schechter formalism is expected to deviate from the -- still not fully known~\cite{Franciolini:2018vbk,Ezquiaga:2018gbw,Cruces:2018cvq,Ezquiaga:2019ftu,Figueroa:2020jkf,Figueroa:2021zah,Hooshangi:2021ubn,Cai:2021zsp,Papanikolaou:2022chm,Animali:2022otk,Pi:2022ysn,Cruces:2022dom,Mishra:2023lhe,Kawaguchi:2023mgk,Hooshangi:2023kss,Stamou:2023vxu,Germani:2023ojx,Raatikainen:2023bzk,Choudhury:2024jlz,Mizuguchi:2024kbl,Escriva:2024aeo} -- true result by many orders of magnitude. In spite of this gap in knowledge, the uncertainties in the GW-spectrum as displayed in~\cref{fig:GWBounds} are manageable, as discussed in~\cref{sec:bounds}.

\subsection{Analytic Approximations for the Power Spectrum}

\begin{figure}[ht]
    \centering
	\captionsetup{font=small}
    \begin{tabular}{cc}
        \includegraphics[width=0.5\textwidth]{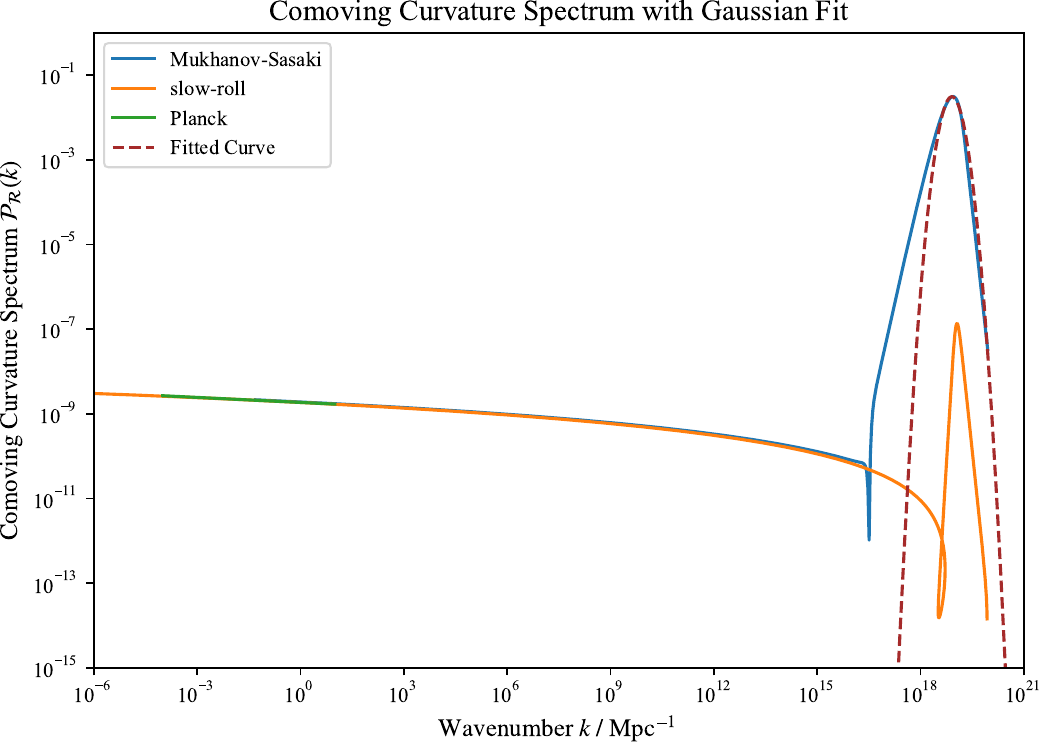} & 
        \includegraphics[width=0.5\textwidth]{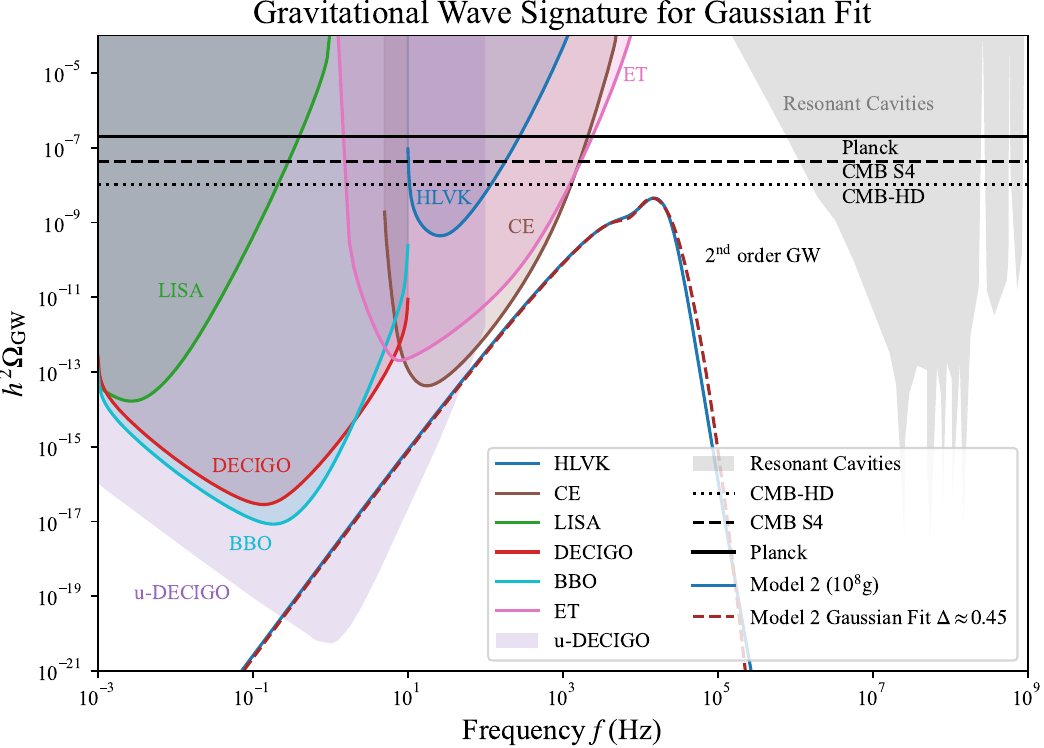} \\
    \end{tabular}
    \caption*{(a) Comparison of fitted power spectrum (left) and GW spectrum (right) for the case with~$\Delta$ free.}

    \vspace{0.5cm} 

    \begin{tabular}{cc}
        \includegraphics[width=0.5\textwidth]{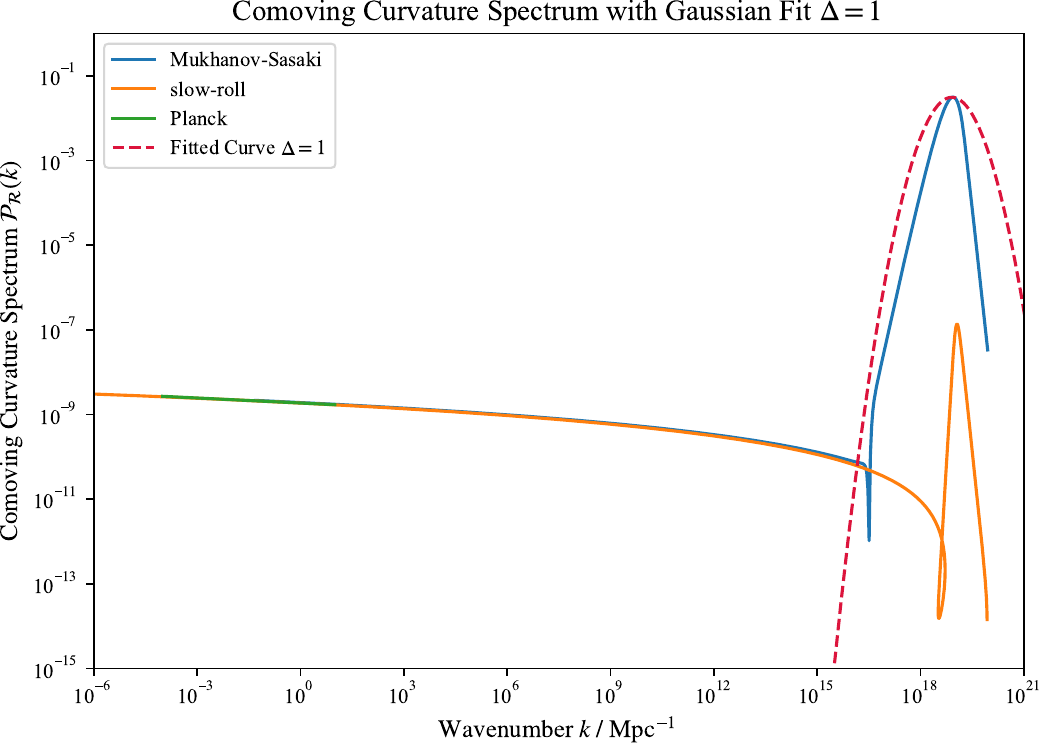} & 
        \includegraphics[width=0.5\textwidth]{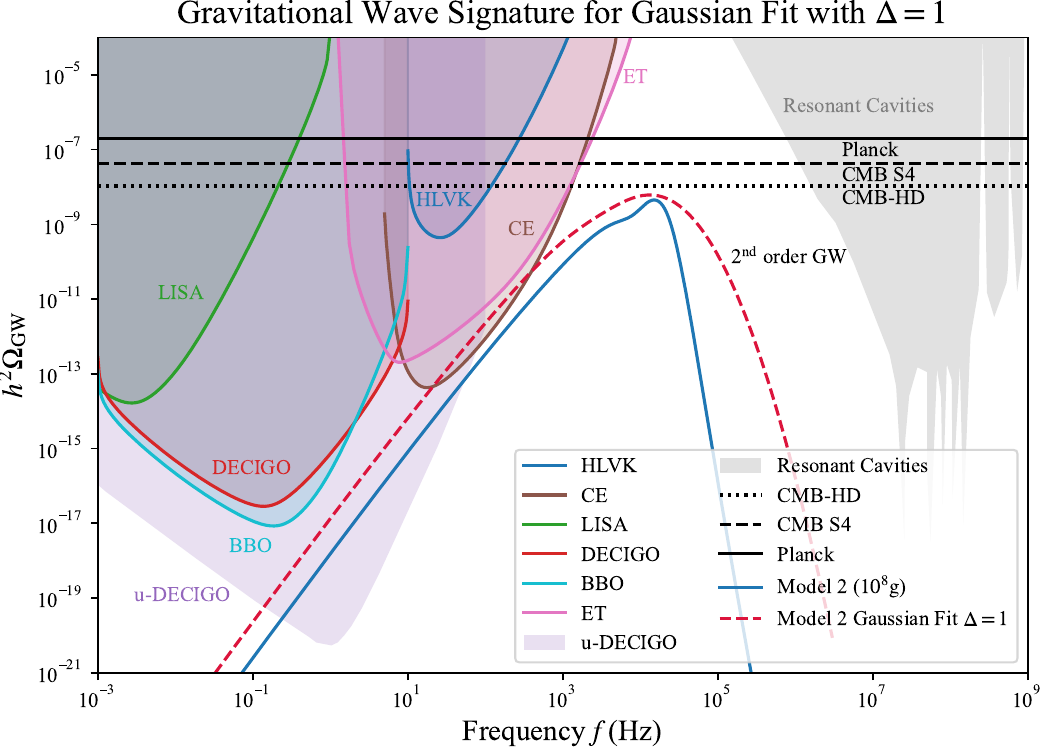} \\
    \end{tabular}
    \caption*{(b) Comparison of fitted power spectrum (left) and GW spectrum (right) for the case with~$\Delta = 1$.}
	\captionsetup{font=normalsize}
    \caption{\label{fig:GaussianComparison} 
    Comparison between the exact power spectrum and the Gaussian fits (left) and their corresponding GW spectrum (right) for model~2. The top row corresponds to~$\Delta$ free, and the bottom row corresponds to~$\Delta = 1$.}
\end{figure}

In previous works~\cite{Franciolini:2023osw,Kohri:2024qpd}, the simultaneous production of small PBHs and scalar-induced GWs has been studied in a different approach. The power spectrum~$\mathcal{P}_\mathcal{R}(k)$ was not derived from inflation, but instead it was modeled with different analytic functions. In this subsection, we will determine to what extent the ans\"atze used in~\cite{Franciolini:2023osw,Kohri:2024qpd} are suited to approximate~$\mathcal{P}_\mathcal{R}(k)$ as derived from our inflationary model. One of the functions employed in~\cite{Kohri:2024qpd} is a Gaussian, which is given by\footnote{In~\cite{Kohri:2024qpd}, the ansatz~$\mathcal{P}_\mathcal{R}(k) = A_{\zeta} \delta(\ln(k/k_0))$ was proposed as second possibility, but we shall not consider this option since such a peaked function is unsuited to describe realistically smooth power spectra, as already discussed in~\cite{Kohri:2024qpd}.}

\begin{equation} \label{GaussianFit}
\mathcal{P}_\mathcal{R}(k) = \frac{A_{\zeta}}{\sqrt{2\pi \Delta^2}} \exp\left(-\frac{(\ln(k/k_0))^2}{2\Delta^2}\right) \;,
\end{equation}
where~$A_{\zeta}$ is the amplitude,~$k_0$ is the peak position, and~$\Delta$ represents the width of the peak. Since in~\cite{Kohri:2024qpd}~$\mathcal{P}_\mathcal{R}(k)$ was not derived from an underlying model, no criterion was available to fix~$\Delta$ and so~$\Delta=1$ was used. In our analysis, we therefore perform two different fits to this Gaussian form: one with~$\Delta$ as a free parameter, and another one with the constraint~$\Delta = 1$. In fixing~$\Delta=1$, we adjust~$A_{\zeta}$ to keep the prefactor~$A_{\zeta}/\sqrt{2\pi \Delta^2}$ the same, which ensures that the PBH abundance does not change significantly. The results of these fits are shown in~\cref{fig:GaussianComparison}. 

Fitting the Gaussian with~$\Delta$ free provides a good approximation around the peak of the power spectrum, with
\begin{equation}
	\Delta \approx 0.45 \;.
\end{equation} 
This leads to nearly identical results for the GW spectrum when compared to the exact calculation. On the other hand, fixing~$\Delta = 1$ results in a poor fit, particularly at the tail of the spectrum, leading to significant deviations in the corresponding GW spectrum. The precise value of~$\Delta$ is therefore critical to the expected GW signal and to assess whether future GW experiments will be sensitive to the signal.

\begin{figure}[ht]
    \centering
    \captionsetup{font=small}
    
    \begin{tabular}{cc}
        \includegraphics[width=0.5\textwidth]{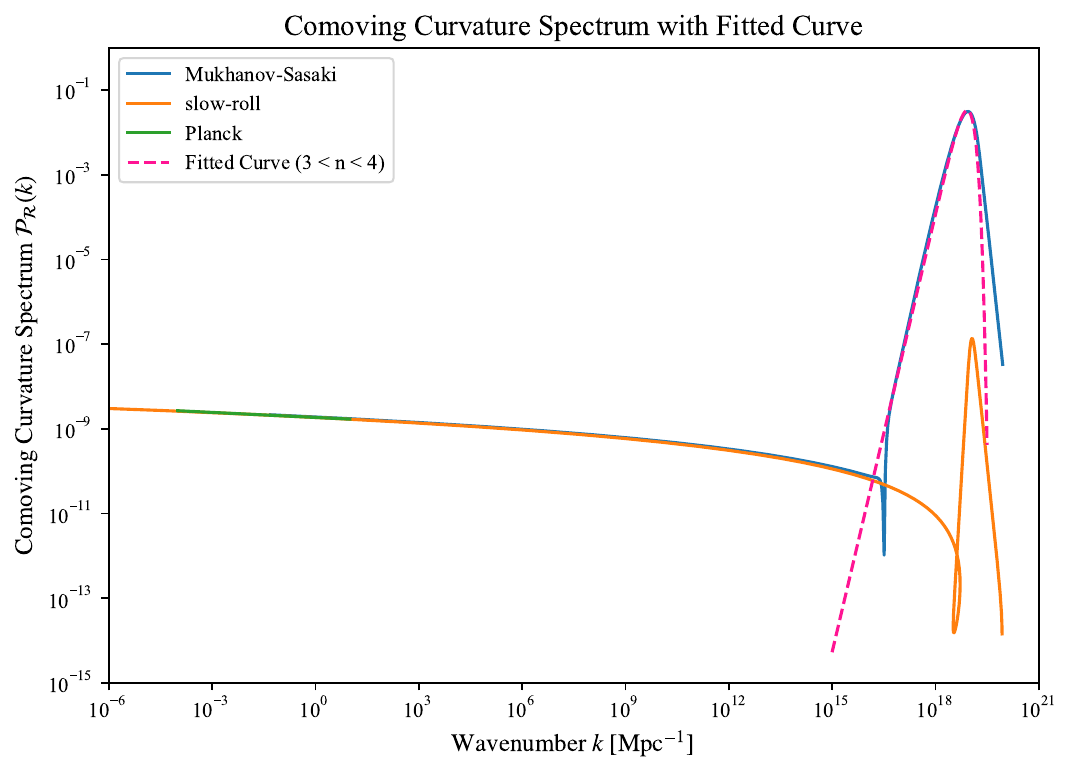} & 
        \includegraphics[width=0.5\textwidth]{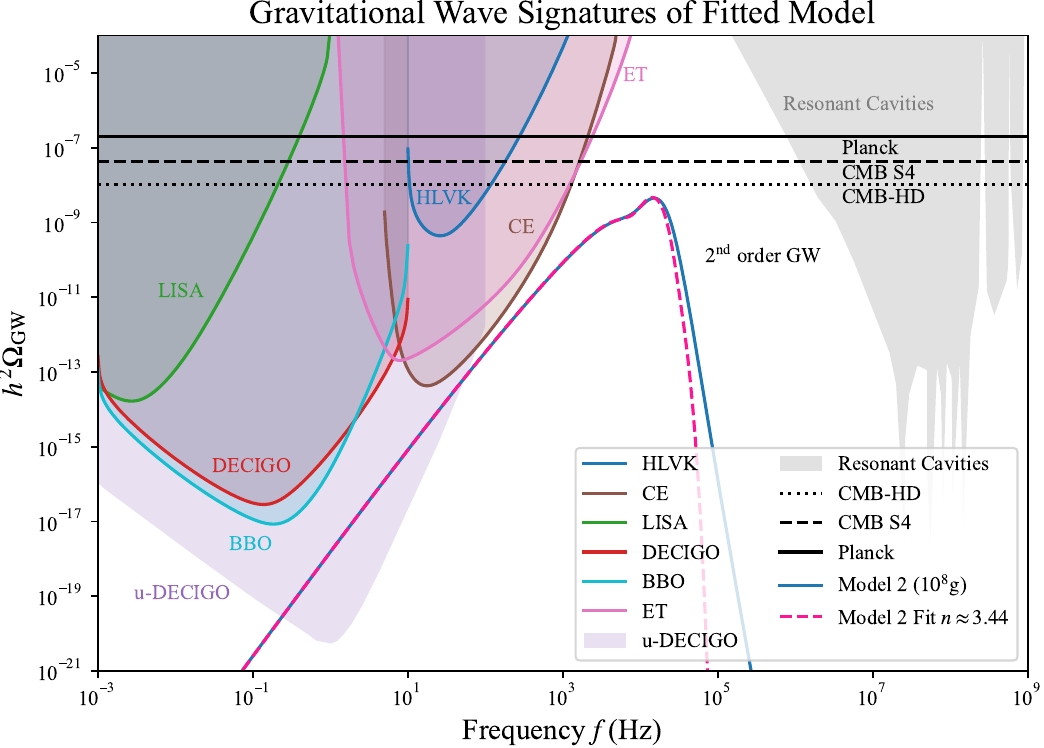} \\
    \end{tabular}
    \caption*{(a) Comparison of fitted power spectrum (left) and GW spectrum (right) for the case with~$n$ free.}

    \vspace{0.5cm} 

    \begin{tabular}{cc}
        \includegraphics[width=0.5\textwidth]{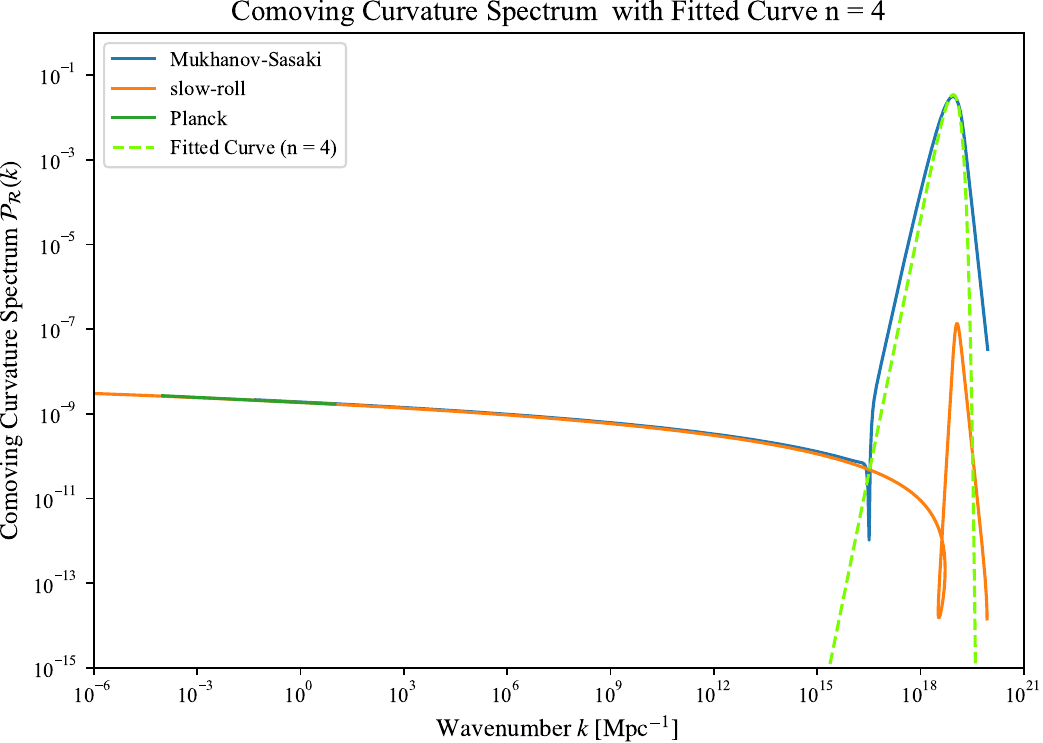} & 
        \includegraphics[width=0.5\textwidth]{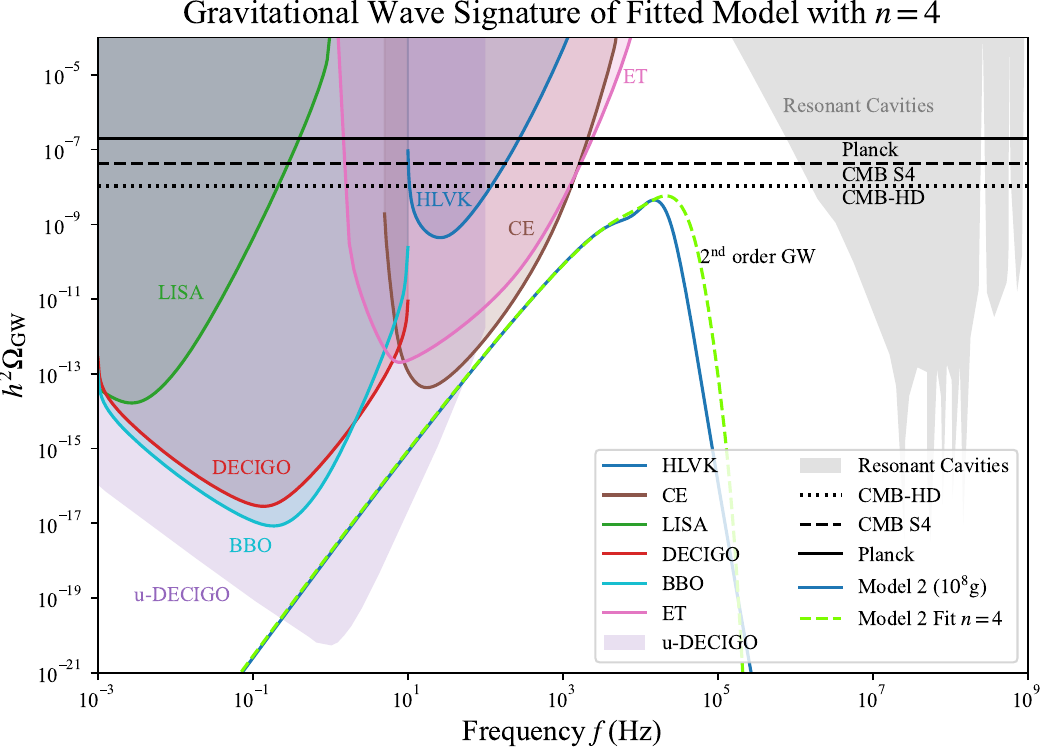} \\
    \end{tabular}
    \caption*{(b) Comparison of fitted power spectrum (left) and GW spectrum (right) for the case with~$n=4$.}
	\captionsetup{font=normalsize}
    \caption{\label{fig:SecondComparison} 
    Comparison between the exact power spectrum and the fit formula proposed in~\cite{Franciolini:2023osw} and their corresponding GW spectrum for model~2.}
\end{figure}

We also test the alternative fit formula proposed in~\cite{Franciolini:2023osw}, which can be written as:
\begin{equation} \label{powerLawFit}
\mathcal{P}_\mathcal{R}(k) = A_{\zeta} \left( \frac{k}{k_0} \right)^n \exp\left[ 2 - 2\left( \frac{k}{k_0} \right)^2 \right] \;,
\end{equation}
where~$A_{\zeta}$ is the peak amplitude,~$n$ is the spectral growth index, and~$k_0$ is the peak position of the power spectrum. This formula attempts to capture both the power-law behavior~$\mathcal{P}_\mathcal{R}\propto k^n$ and the exponential suppression of modes beyond the peak. Since~$n=4$ was used in~\cite{Franciolini:2023osw}, we will again perform two fits, one with~$n$ free and another one with the constraint~$n=4$. We show the result in~\cref{fig:SecondComparison}. Fitting with~$n$ free,~\cref{powerLawFit} provides a good fit at low wavenumbers with
\begin{equation}
	n \approx 3.44 \;,
\end{equation}
but deviates soon beyond the peak. This can be seen in the GW spectra, with only minor deviations from the exact result to the right of the peak. For fixed~$n=4$, the fit to the power spectrum is visibly worse and the GW spectrum has a larger peak than in actuality.

We conclude that both functions~\cref{GaussianFit} and~\cref{powerLawFit} are well-suited to approximate the peak of the power spectrum and the resulting GW signal. However, the choice of parameters in the approximations for the spectra is critical to the determining possibility of detection with future GW experiments. For example, using the Gaussian approximation~\cref{GaussianFit} and~$\Delta=1$ for the power spectrum,~\cite{Kohri:2024qpd} identifies a detection possibility with CE already for PBHs with masses greater than~$10^7\,\mathrm{g}$. However, our results indicate that PBHs of greater mass, around~$10^8\,\mathrm{g}$, are required due to the reduced width of the peak in the power spectrum affecting the shape of the GW signal.

\section{Conclusion} \label{sec:conclusion}
We know astonishingly little about what happens when black holes lose a sizable fraction of their mass. It is possible that evaporation slows down significantly so that small primordial black holes (PBHs) below~$\sim10^{14}\,\mathrm{g}$ become viable dark matter candidates~\cite{Dvali:2020wft}. Whether such a slowdown is indeed realized in Nature remains to be determined, and arguably one of the best ways of doing so is to derive observable implications. In this spirit, we have investigated scalar-induced gravitational waves (GW) that result from the inflationary production of long-lived small PBHs. We assume that the PBHs comprise a substantial fraction of the dark matter of our Universe and following the results of~\cite{Alexandre:2024nuo,Thoss:2024hsr} we focus on the mass range~$10^6-10^9\,\mathrm{g}$. We conclude with a list of our findings:
\begin{itemize}
    \item In a variation of the model~\cite{Ballesteros:2020qam}, we consider a scenario of single-field inflation with a renormalizable polynomial potential and a non-minimal coupling~$\xi\sqrt{-g}\phi^2R$ between the inflaton and the Ricci scalar. We show that a near inflection point a few e-folds before the end of inflation allows for the formation of a significant abundance of small PBHs that can account for the entirety of dark matter.
    \item The model is in full agreement with the Planck~\cite{Planck:2018jri} and BICEP~\cite{BICEP:2021xfz} observations of the CMB. Considering smaller mass PBH resolves the~$\sim\!3\sigma$ tensions between the predicted spectral index~$n_s$ and the Plank measurements found with models producing PBHs with masses above the Hawking evaporation limit~\cite{Ballesteros:2020qam}. The predicted spectral index is at the lower end of observational interval, so future CMB observations can test this scenario.
    \item Although large theoretical uncertainties plague expectations of PBH abundance, we can constrain the accompanying stochastic GW background of models that allow PBH to account for the entirety of dark matter to within a few orders of magnitude. Moreover, the GW signal is largely independent of the time at which the semi-classical description of a black hole breaks down and so also applies to the case of an early slowdown~\cite{Michel:2023ydf}.
    \item The peak in the GW signal falls in a frequency region that appears difficult to constrain with current proposals for GW detectors. Still, there can be a  potential for detection by CMB-HD and an ideal configuration of u-DECIGO, and moreover models that produce PBHs above~$\sim 10^8\,\mathrm{g}$ may be constrained by Cosmic Explorer and Einstein Telescope. Our results clearly indicate the need to develop experimental approaches for GW detection around~$10^5\mathrm{Hz}$.
    \item We examined two analytic approximations for the power spectrum~\cite{Franciolini:2023osw, Kohri:2024qpd} and showed that the resulting GW signatures are highly sensitive to the choice of parameters, such as the peak width~$\Delta$ and the spectral index~$n$, with these differences producing notable variations in the GW spectrum.
\end{itemize}
The stochastic GW background is an exciting avenue for exploring inflationary models that produce PBHs as the primary constituent of dark matter.  Furthermore, detection of a single long-lived small PBH would validate the idea that the semi-classical description of a BH breaks down and that a slow down of evaporation occurs. This would greatly enhance our understanding of how BHs evolve and the principles of quantum gravity.

\begin{acknowledgments}
	We are grateful to Gia Dvali, Gabriele Franciolini and Yannis
	Georis for discussions and insightful feedback, and we are indebted to Luis Anchordoqui for making us aware of an incorrect statement regarding mass ranges of PBHs in a previous version. We thank Ana Alexandre, Guillermo Ballesteros, Takahiro Terada and Valentin Thoss for correspondence as well as Yannis Georis for making the sensitivity data of~\cite{Drewes:2023oxg} available to us.
	
	This work used the DiRAC Data Intensive service~(CSD3 \href{www.csd3.cam.ac.uk}{www.csd3.cam.ac.uk}) at the University of Cambridge, managed by the University of Cambridge University Information Services on behalf of the STFC DiRAC HPC Facility~(\href{www.dirac.ac.uk}{www.dirac.ac.uk}). The DiRAC component of CSD3 at Cambridge was funded by BEIS, UKRI and STFC capital funding and STFC operations grants. DiRAC is part of the UKRI Digital Research Infrastructure.
This work also used the Newton server, access to which was provisioned by Will Handley using an ERC grant.

WB is grateful for the support of Girton College, Cambridge, Marie Skłodowska-Curie Actions and the Institute of Physics of the Czech Academy of Sciences. SZ acknowledges support by the European Research Council Gravites Horizon Grant AO number: 850 173-6 and the \emph{Fonds de la Recherche Scientifique} -- FNRS.

\textbf{Disclaimer:} Funded by the European Union. Views and opinions expressed are however those of
the authors only and do not necessarily reflect those of the European Union or European Research
Council. Neither the European Union nor the granting authority can be held responsible for them.
\end{acknowledgments}
	
	\appendix
	
	\section{Background of inflation, PBHs and GWs}
	\label{app:background}
	\textbf{Convention:} In this appendix we will additionally choose~$M_P\equiv1$.
	\subsection{Generation of large inflationary perturbations}\label{app:perturbations}
    We consider a canonically normalised inflaton field~$h$ described by the action~\cref{canonicalInflaton}.
	In the spatially flat Friedmann–Lema\^{\i}tre–Robertson–Walker metric with scale factor~$a(t)$, and the number of e-folds of inflation between initial cosmic time~$t_i$ and cosmic time~$t$ given by
	\begin{equation}
		N_e(t)=\int_{t_i}^tH(t')\mathrm{d}t' \;,
	\end{equation}
	for~$H(t)\equiv a^{-1}\mathrm{d}a/\mathrm{d}t$ ~\cite{Hobson:2006se,Baumann:2022mni}. The dynamics of~$h$ are determined by the equation of motion~\cite{Ballesteros:2017fsr,Ballesteros:2020qam},
	\begin{equation}\label{eom}
		\frac{\mathrm{d}^2h}{\mathrm{d}N_e^2}+3\frac{\mathrm{d}h}{\mathrm{d}N_e}-\frac{1}{2}\left(\frac{\mathrm{d}h}{\mathrm{d}N_e}\right)^3+\left[3-\frac{1}{2}\left(\frac{\mathrm{d}h}{\mathrm{d}N_e}\right)^2\right]\frac{\mathrm{d}\log{U}}{\mathrm{d}h}=0 \;,
	\end{equation}
	with slow roll (SR) initial conditions~\cite{Liddle:1994dx, Ballesteros:2014yva}.~\cref{eom} is exact and is solved numerically until the end of inflation, defined to be when~$\epsilon_H=1$ for\footnote{Leading up to the phase of USR~$\epsilon_H$ may grow to briefly exceed 1, but it becomes small again during the USR phase. Inflation is defined to end at~$\epsilon_H=1$ after the USR phase.}
	\begin{equation}
		\epsilon_H \equiv-\frac{\dot{H}}{H^2}=\frac{1}{2}\left(\frac{\mathrm{d}h}{\mathrm{d}N_e}\right)^2 \;.
	\end{equation}
	
	$V(\phi)$ is chosen to have a stationary inflection point so that the inflaton velocity slows down so that it just makes it over the barrier, see~\cref{fig:inflatonandpotential} in the main text. This produces a phase of ultra slow roll (USR) evolution, formally defined by~$\eta_H>3$ for
	\begin{equation}\label{eta}
		\eta_H\equiv-\frac{\ddot{H}}{2H\dot{H}}=\epsilon_H-\frac{1}{2}\frac{\mathrm{d}\log{\epsilon_H}}{\mathrm{d}N_e} \;.
	\end{equation}
	The phase of USR boosts the comoving perturbations produced at large wavenumbers. Nevertheless, the non-minimal coupling to gravity ensures the potential is sufficiently flat such that \emph{at CMB scales} the comoving curvature perturbation spectrum is well described by the SR approximation~\cite{Ballesteros:2017fsr}. In this limit the comoving curvature perturbation spectrum spectrum becomes~\cite{Stewart:1993bc,Baumann:2022mni}
	\begin{equation}\label{SRPower}
		\mathcal{P}_\mathcal{R}(k)=\frac{H^2}{8\pi^2\epsilon_H}\left(\frac{k}{aH}\right)^{-4\epsilon_H+2\eta_H}=A_s\left(\frac{k}{aH}\right)^{n_s-1} \;,
	\end{equation}
	with spectral index
	\begin{equation}\label{spectralindex}
		n_s=1-4\epsilon_H+2\eta_H \;,
	\end{equation}
	and scalar amplitude~$A_s=H^2/(8\pi^2\epsilon_H)$. The tensor-to-scalar ratio becomes
	\begin{equation}\label{tensorscalar}
		r\equiv\frac{A_\mathrm{t}}{A_\mathrm{s}}=16\epsilon_H \;,
	\end{equation}
	where~$A_\mathrm{t}$ is the amplitude of the tensor power spectrum. These quantities are evaluated at the horizon exit of the pivot scale~$k=k_*=0.05\mathrm{Mpc}^{-1}$.
	
	However, at larger wavenumbers the inflaton stalls at inflection point causing a phase of USR that boosts the comoving perturbation spectrum and for significant PBH formation. This breaks the assumptions of SR~\cite{Pattison:2018bct, Motohashi:2017kbs} and a more precise calculation of the power spectrum using the Mukhanov-Sasaki formalism~\cite{Sasaki:1986hm, Mukhanov:1988jd} is required to accurately account for the perturbations~\cite{Ballesteros:2017fsr,Chongchitnan:2006wx,Motohashi:2017kbs,Germani:2017bcs}. In Fourier space, and in terms of conformal time ($a\mathrm{d}\tau=\mathrm{d}t$), the Mukhankov-Sasaki equation is
	\begin{equation}\label{MS}
		\frac{\partial^2 f_k}{\partial \tau^2}+\left( k^2-\frac{1}{z}\frac{\mathrm{d}^2z}{\mathrm{d}\tau^2}\right)f_k=0 \;,
	\end{equation}
	where~$z$ is defined by
	\begin{equation}
	z\equiv\frac{1}{H}\frac{\mathrm{d}h}{\mathrm{d}\tau} \;.
	\end{equation}
	The function~$f_k$ is related to the comoving curvature perturbation~$\mathcal{R}$ by~$\mathcal{R}=-f_k/z$. The mode function~$f_k$ satisfies the initial condition~\cite{Baumann:2022mni,Ballesteros:2017fsr}
	\begin{equation}
		\lim_{k\tau\to-\infty} f_k(\tau)=\frac{1}{\sqrt{2k}}e^{-ik\tau}
	\end{equation}
	in the formal limit when~$\tau\to-\infty$, \ie when the mode is deep inside the Hubble radius,~$k\gg aH$. The comoving curvature perturbation spectrum for~$\mathcal{P}_\mathcal{R}(k)$ is evaluated at superhorizon scales $k\ll aH$ where~$\mathcal{R}$ is frozen
	\begin{equation}\label{powerspectrum}
		\mathcal{P}_\mathcal{R}(k)=\frac{k^3}{2\pi^2}\biggl|\frac{f_k}{z}\biggl|^2_{k\ll aH} \;.
	\end{equation}
	It is computationally more efficient to solve the Mukhanov-Sasaki equation using e-folds~$N_e$ instead of proper time~$\tau$. When expressed in terms of~$N_e$, the Mukhankov-Sasaki equation for the Fourier mode~$f_k$ becomes
	\begin{equation}
		\frac{\mathrm{d}^2f_k}{\mathrm{d}N_e^2}+(1-\epsilon_H)\frac{\mathrm{d}f_k}{\mathrm{d}N_e}+\left[\frac{k^2}{a^2H^2}+(1+\epsilon_H-\eta_H)(\eta_H-2)-\frac{\mathrm{d}}{\mathrm{d}N_e}(\epsilon_H-\eta_H)\right]f_k=0 \;,
	\end{equation}
	with the boundary conditions
	\begin{subequations}
		\begin{equation}
			f_k=\frac{1}{\sqrt{2k}} \;,
		\end{equation}
		\begin{equation}
			\frac{\mathrm{d}f_k}{\mathrm{d}N_e}=\frac{-i\sqrt{k}}{\sqrt{2}k_i} \;,
		\end{equation}
	\end{subequations}
	where~$k_i(\ll k)$ is the wavenumber leaving the Hubble horizon at the start of the integration. We found it sufficient to choose~$k_i$, to be 20 times smaller than~$k$. Each mode was evolved until it becomes frozen ($\sim\!10$ e-folds after Hubble crossing is found to be sufficient) where the frozen comoving curvature perturbation is evaluated.
	
	\subsection{Production of primordial black holes}\label{app:pbh_production}
	The mass of a PBH is directly proportional to the mass inside the Hubble horizon upon horizon re-entry during the radiation-dominated era~\cite{Sasaki:2018dmp}. For relating the mass of the PBH and the wavenumber of the curvature perturbation, the following relation was used in \cite{Ballesteros:2017fsr,Ballesteros:2020qam} (see \cite{Carr:1975qj}):
	\begin{equation}\label{mass}
		M(k)\simeq 10^{17}\left(\frac{k}{2\times10^{14}\mathrm{Mpc^{-1}}}\right)^{-2}\,\mathrm{g}\;.
    \end{equation}	
	This results in a peak in the GW signal at a frequency
	\begin{equation} \label{frequency}
		f_\mathrm{peak} \simeq 977 \left(\frac{M}{10^{10}\, 	\mathrm{g}}\right)^{-1/2}\,\mathrm{Hz} \;,
\end{equation}
which is in agreement with~\cite{Kohri:2024qpd}. While additional corrections to $M(k)$ have been derived (see \cite{Franciolini:2023osw} and references therein), we shall employ \eqs \eqref{mass} and \eqref{frequency} in the following in order to allow for an easier comparison with \cite{Ballesteros:2020qam,Kohri:2024qpd}.\footnote{In \cite{Franciolini:2023osw}, $M(k)\simeq 9.02\times10^{46}\left(k/ \mathrm{Mpc^{-1}}\right)^{-2}\,\mathrm{g}$ was used instead of our \cref{mass},~$M(k)\simeq 4\times10^{45}\left(k/ \mathrm{Mpc^{-1}}\right)^{-2}\,\mathrm{g}$. In terms of peak GW frequency~$f_\mathrm{peak}$, this results in a shift by a factor of approximately~$4.7$. In other words, our results such as \fig \ref{fig:GWs} remain valid also with the relation of~\cite{Franciolini:2023osw}, provided that the PBHs mass associated to each GW curve is multiplied by a factor of~$\sim 22$.}
	
	In the Press–Schechter formalism~\cite{Press:1973iz,Ballesteros:2017fsr} of gravitational collapse, and assuming Gaussian fluctuations, the mass fraction of dark matter in PBHs of mass~$M$, is given by the probability of an overdensity~$\delta$ being above a critical threshold~$\delta_c$. The standard value for the threshold is~$\delta_c=0.45$~\cite{Green:2004wb, Musco:2004ak} but recent numerical models indicate~$\delta_c$ could be anywhere in the range~$0.4-0.6$~\cite{Musco:2004ak, Harada:2015yda}. The fraction of dark matter in the form of PBHs of mass~$M$ is~\cite{Ballesteros:2020qam}
	\begin{align}\label{fPBH}
		f_\mathrm{PBH}&=\frac{\Omega_\mathrm{PBH}}{\Omega_\mathrm{DM}}\\&\simeq \left(\frac{M}{10^{18}\,\mathrm{g}}\right)^{-1/2}\frac{1.25\times10^{15}}{\sqrt{2\pi\sigma^2(M)}}\int^\infty_{\delta_c} \exp{\left[-\frac{\delta^2}{2\sigma^2(M)}\right]}\mathrm{d}\delta \;.
	\end{align}
	which is determined by the variance~$\sigma^2(M)$ assumed to be the coarse-grained 
	variance of the density contrast smoothed on a scale~$R=1/k$ given by~\cite{1403.1198}
	\begin{equation}
		\sigma^2(M)=\frac{16}{81}\int^\infty_0\frac{\mathrm{d}q}{q}\left(\frac{q}{k}\right)^4\mathcal{P}_\mathcal{R}(q)W^2\left(q/k\right) \;.
	\end{equation}
	There is no unique prescription for the window function, common choices include a Gaussian~$W(x)=\exp{(-x^2/2)}$, or a real-space top-hat function~$W(x)=3(\sin{x}-x\cos{x})/x^3$~\cite{Ando:2018qdb}. The PBH fraction can vary by over 5 orders of magnitude depending on the choice of window function~\cite{Dandoy:2023jot}. In previous works (e.g.,~\cite{Ando:2018qdb,Kohri:2024qpd}), a transfer function has been included in the calculation of the PBH abundance. In our calculation, we have opted to omit it, as its effect is much smaller than the other uncertainties in the Press-Schechter approach and therefore it does not significantly impact our results.

	\subsection{Secondary gravitational waves}\label{app:gwcalc}
	Large primordial curvature perturbations generated during inflation, whether responsible or not for PBHs as dark matter, inevitably act as a second-order source of primordial GWs~\cite{Espinosa:2018eve,Kohri:2018awv, Domenech:2021ztg}. The GW signature may be computed directly from the comoving curvature perturbation spectrum~$\mathcal{P}_\mathcal{R}(k)$. In the following, we emphasize the crucial points for the computation, directing interested readers to~\cite{Espinosa:2018eve} for comprehensive details. The spectral energy density of GW, measured relative to the critical energy density today, is expressed as
	\begin{equation}
		\Omega_\mathrm{GW}(k)=\frac{c_g\Omega_r}{36}\int_0^{\frac{1}{\sqrt{3}}} \mathrm{d}t\int^\infty_{\frac{1}{\sqrt{3}}} \mathrm{d}sT(s,t)\mathcal{P}_\mathcal{R}(ku)\mathcal{P}_\mathcal{R}(kv) \;,
	\end{equation}
	which is conventionally plotted as a function of frequency using~$f=k/2\pi$ (becoming~$f=1.546\times10^{-15}[k/1\mathrm{Mpc}^{-1}]\,\mathrm{Hz}$ when dimensionality is restored). We have
	\begin{equation}
		T(s,t)\equiv\bigg[\frac{(t^2-1/3)(s^2-1/3)}{t^2-s^2}\bigg]^2\left[\mathcal{I}_c(t,s)^2+\mathcal{I}_s(t,s)^2\right] \;,
	\end{equation}
	and~$\Omega_r$ is the current energy density of radiation,~$u\equiv\sqrt{3}(s+t)/2$,~$v\equiv\sqrt{3}(s-t)/2$, and~$\mathcal{I}_c$ and~$\mathcal{I}_s$ are two analytic functions
	\begin{equation}
		\mathcal{I}_c(t,s) = -36\pi\frac{s^2+t^2-2)^2}{(s^2-t^2)^3}\theta(s-1) \;,
	\end{equation}
	\begin{equation}
		\begin{split}
			\mathcal{I}_s(t,s)=-36\frac{(s^2+t^2-2)}{(s^2-t^2)^2}\bigg[\frac{(s^2+t^2-2)}{(s^2-t^2)}\log{\frac{(1-t^2)}{|s^2-1|}}+2\bigg] \;,
		\end{split}
	\end{equation}
	for~$\theta(x)$ the unit step function. The parameter~$c_g$ accounts for the change of effective degrees of freedom during the QCD phase transition. It is approximately a constant~$c_g\approx0.4$ at the frequencies considered~\cite{Frosina:2023nxu}.

    \section{Details about numerical procedure}\label{app:numerical}
    \subsection{Numerical instabilities}\label{app:instabilities}
	Various numerical instabilities can occur in the computations described in~\cref{sec:numerical}. The key issues that arose, and the manner in which they were resolved, are detailed below.
    \begin{itemize}
    \item \emph{Solving for the background dynamics becomes unstable and diverges}\vspace{0.1cm}\\
    This occurs if the inflaton becomes stuck in the local minimum of the potential, and if not then shortly after the end of inflation. This is due to stiffness in the equations. These divergences are not inherently a problem provided we can identify that the solution either became stuck or that inflation has ended.
    \item \emph{Evolution of the Mukhanov-Sasaki modes can be unstable}\vspace{0.1cm}\\
	   Quite apart from the selection of an efficient integration algorithm, we have three more rudimentary ways to help mitigate this instability. Firstly, we solve the real and imaginary parts separately. Secondly, we scale the initial conditions down (typically by a factor of~$10^6$), evolve the equations, and then scale the solution back up again. This improves stability because~$f_k$ grows rapidly after the mode exits the horizon; the growth of~$f_k$ can cause the integration to become stiff. Finally, we control how far outside of the horizon the modes are evolved. We must be careful to ensure that the mode evolves enough to become sufficiently close to its frozen value. Evolving too far, however, causes~$f_k$ to grow too large and the integration to again become stiff. Despite these difficulties, fast integration of the stiff Mukhanov-Sasaki equation is found to be well within the capabilities of implicit Runge-Kutta methods~\cite{Axelsson1969ACO,Ehle:1973} (see more general discussions in~\cite{Butcher:1996,Hairer:1996}).\footnote{Specifically, we employ the \texttt{solve\_ivp} implementation in the SciPy Python library~\cite{2020SciPy-NMeth}, where the optional argument \texttt{method='Radau'} selects Axelsson's type-IIA method, itself based on Radau quadrature. After some experimentation with the available libraries, this combination was found to be optimally stable, fast and accurate when applied to the unusually challenging Mukhanov-Sasaki integration in this paper (see supplemental materials~\cite{Supplement}).}
    \item \emph{PBH abundance calculation may not converge}\vspace{0.1cm}\\
    The PBH abundance is very sharply peaked and so to ensure convergence we need to find the peak using ~\cref{mass} and integrate in a small mass interval around it.
    \end{itemize}

    \subsection{Procedural Tuning of~$\xi$}\label{app:tuning_xi}
    Increasing~$\xi$ acts to increase~$\mathcal{P}^\mathrm{peak}_\mathcal{R}$, but increasing~$\xi$ too far results in the inflaton becoming stuck at the local minimum in the potential. To boost~$\mathcal{P}^\mathrm{peak}_\mathcal{R}$ we must fine-tune~$\xi$ to remain just below the critical value at which the inflaton becomes stuck~$\xi^\mathrm{critical}$, typically to within~$\frac{\xi^\mathrm{critical}-\xi}{\xi^\mathrm{critical}}\lesssim10^{-6}$. To identify values of~$\xi$ that give significant boosting of~$\mathcal{P}^\mathrm{peak}_\mathcal{R}$, we use the following algorithm:
    \begin{enumerate}
        \item Choose a value of~$\xi$ that is known to cause the inflaton to get stuck at the inflection point,~$\xi_\mathrm{big}$, and a value that does not~$\xi_\mathrm{small}$ (typically we may choose 1 and 0 respectively).
        \item Evaluate the background evolution with the values~$\xi=\xi_\mathrm{avg.}=\frac{\xi_\mathrm{big}+\xi_\mathrm{small}}{2}$ and~$\lambda=1\times10^{-9}$ ($\lambda$ is chosen arbitrarily and is fine-tuned after a value of~$\xi$ is chosen).
        \item If the inflaton gets stuck at the inflection point, set~$\xi_\mathrm{big}$ to~$\xi_\mathrm{avg.}$. If not, find~$\mathcal{P}^\mathrm{peak}_\mathcal{R}$ and set~$\xi_\mathrm{small}$ to~$\xi_\mathrm{avg.}$.
        \item If the desired boost in the power spectrum has not been reached, go to step 2.
    \end{enumerate}
    Since the evaluation of the background dynamics can be time-consuming, it is best to record the maximum power at each value of~$\xi$ where the inflaton did not get stuck~$\mathcal{P}^\mathrm{peak}_\mathcal{R}(\xi)$. Adjusting the value of the peak in the power spectrum can then be achieved by interpolating between these values.
    
    The power spectrum is computed in the SR approximation using~\cref{SRPower} since the Mukhanov-Sasaki approach can be unstable (and often needs human oversight). Nevertheless, since Mukhanov-Sasaki power spectrum is always around a factor of~${\sim}10^{1}-10^2$ larger than the approximate power spectrum, this approach can also be used to adjust the peak value in the Mukhanov-Sasaki power spectrum and hence the fractional abundance of PBH as dark matter.
 
	\bibliographystyle{JHEP}
	\bibliography{RefsInspire,RefsOther}
	
\end{document}